\documentclass[twocolumn]{aastex631}

\newcommand\Msun{$M_{\odot}$} % \newcommand\Msun{$\mathrm{M_{\odot}}$}
\newcommand\Rsun{$R_{\odot}$} % \newcommand\Rsun{$\mathrm{R_{\odot}}$}
\newcommand\MJ{$M_{\rm J}$} %\newcommand\MJ{$\mathrm{M_{\rm J}}$}
\newcommand\RJ{$R_{\rm J}$} %\newcommand\RJ{$\mathrm{R_{\rm J}}$}
\newcommand\RE{$R_{\Earth}$} % \newcommand\RE{$\mathrm{R_{\Earth}}$}
\newcommand\ME{$M_{\oplus}$} % \newcommand\ME{$\mathrm{M_{\oplus}}$}
\newcommand\ab{mag$_\mathrm{{AB}}$}

\usepackage{multirow}
\shorttitle{The TEMPO Survey I}
\shortauthors{Limbach et al.}
\graphicspath{{./}{figures/}}

\begin{document}https://www.overleaf.com/project/61c9de1cf011dc621f745594
\title{The TEMPO Survey I: Predicting Yields of the Transiting Exosatellites, Moons, and Planets from a 30-day Survey of Orion with the Nancy Grace Roman Space Telescope}

\correspondingauthor{Mary Anne Limbach}
\email{maryannelimbach@gmail.com}

\author[0000-0002-9521-9798]{Mary Anne Limbach}
\affiliation{Department of Physics and Astronomy, Texas A\&M University, 4242 TAMU, College Station, TX 77843-4242 USA}

\author[0000-0001-7493-7419]{Melinda Soares-Furtado}
\altaffiliation{NASA Hubble Science Fellow}
\affiliation{Department of Astronomy,  University of Wisconsin-Madison, 475 N.~Charter St., Madison, WI 53703, USA}

\author[0000-0001-7246-5438]{Andrew Vanderburg}
\affiliation{Department of Physics and Kavli Institute for Astrophysics and Space Research, Massachusetts Institute of Technology, Cambridge, MA 02139, USA}

\author[0000-0003-0562-1511]{William M. J. Best}
\affil{Department of Astronomy, The University of Texas at Austin, 2515 Speedway C1400, Austin, TX 78712, USA}

\author[0000-0002-3656-6706]{Ann Marie Cody}
\affil{SETI Institute, 189 N Bernardo Ave \#200, Mountain View, CA 94043 USA
}

\author[0000-0003-2676-8344]{Elena D'Onghia}
\affiliation{Department of Astronomy,  University of Wisconsin-Madison, 475 N.~Charter St., Madison, WI 53703, USA}

\author[0000-0002-9831-0984]{Ren\'{e} Heller}
\affil{Max Planck Institute for Solar System Research, Justus-von-Liebig-Weg 3, 37077 G\"{o}ttingen, Germany}

\author[0000-0001-7449-4638]{Brandon S.~Hensley}
\affil{Department of Astrophysical Sciences, Peyton Hall, Princeton University, Princeton, NJ, USA 08544}

\author[0000-0002-5365-1267]{Marina Kounkel}
\affil{Department of Physics and Astronomy, Vanderbilt University, Nashville, TN 37235, USA}

\author[0000-0001-9811-568X]{Adam Kraus}
\affil{Department of Astronomy, The University of Texas at Austin, 2515 Speedway C1400, Austin, TX 78712, USA}

\author[0000-0002-4128-6901]{Andrew W.~Mann}
\affil{Department of Physics and Astronomy, The University of North Carolina at Chapel Hill, Chapel Hill, NC 27599-3255, USA}
%\affil{Harlan J.~Smith Fellow}

\author[0000-0002-9573-3199]{Massimo Robberto}
\affil{Space Telescope Science Institute, 3700 San Martin Drive, Baltimore, MD 21218, USA}
\affil{Johns Hopkins University, 3400 N.~Charles Street, Baltimore, MD 21218, USA}

\author[0000-0003-4423-0660]{Anna L.~Rosen}
\affil{Center for Astrophysics and Space Sciences, University of California, San Diego, 9500 Gilman Drive, La Jolla, CA 92093, USA}

\author[0000-0002-2522-8605]{Richard Townsend}
\affiliation{Department of Astronomy,  University of Wisconsin-Madison, 475 N.~Charter St., Madison, WI 53703, USA}

\author[0000-0003-0489-1528]{Johanna M.~Vos}
\affil{Department of Astrophysics, American Museum of Natural History, Central Park West at 79th Street, New York, NY 10024, USA}

\begin{abstract}
We present design considerations for the Transiting Exosatellites, Moons, and Planets in Orion (TEMPO) Survey with the Nancy Grace Roman Space Telescope.
This proposed 30-day survey is designed to detect a population of transiting extrasolar satellites, moons, and planets in the Orion Nebula Cluster (ONC).
The young (1--3\,Myr), densely-populated ONC harbors about a thousand bright brown dwarfs (BDs) and free-floating planetary-mass objects (FFPs). 
TEMPO offers sufficient photometric precision to monitor FFPs with M~$\geq1$\,\MJ{} for transiting satellites.
The survey is also capable of detecting FFPs down to sub-Saturn masses via direct imaging, although follow-up confirmation will be challenging. 
TEMPO yield estimates include 14 (3-22) exomoons/satellites transiting FFPs and 54 (8-100) satellites transiting BDs. Of this population, approximately $50\%$ of companions would be ``super-Titans" (Titan to Earth mass). 
Yield estimates also include approximately $150$ exoplanets transiting young Orion stars, of which $>50\%$ will orbit mid-to-late M dwarfs and approximately ten will be proto-habitable zone, terrestrial (0.1\,\ME{} - 5\,\ME{}) exoplanets. 
TEMPO would provide the first census demographics of small exosatellites orbiting FFPs and BDs, while simultaneously offering insights into exoplanet evolution at the earliest stages. 
This detected exosatellite population is likely to be markedly different from the current census of exoplanets with similar masses (e.g., Earth-mass exosatellites that still possess H/He envelopes). Although our yield estimates are highly uncertain, as there are no known exoplanets or exomoons analogous to these satellites, the TEMPO survey would test the prevailing theories of exosatellite formation and evolution, which limit the certainty surrounding detection yields. 
\end{abstract}


\section{Introduction} \label{sec:intro}
The discovery of companions via the transit method typically requires long-term monitoring of numerous sources within a large field of view (FOV). Here we use the term ``companion" to refer to a secondary object orbiting a primary of any mass, including stellar-mass hosts, brown dwarfs (BDs), and young free-floating planetary-mass objects (FFPs)\footnote{The naming convention used in this manuscript was chosen to be consistent with the Roman Microlensing Survey \citep{2020AJ....160..123J}.}; these latter sources are also known as ``isolated planetary-mass objects," or, when ejected from their stellar hosts, ``rogue planets". Following a precedent established in the literature, we use the term ``exosatellite" or ``satellite" to refer to a companion orbiting an FFP, BD, or late M dwarf \citep{2015ApJ...800..126K,2019BAAS...51c.169M,2022AJ....163..253T}. We use a mass cutoff of 13\,\MJ{} to differentiate between FFP and BD \citep{2011ApJ...727...57S}. 

The transit technique continues to play a critical role in exoplanet detection, unveiling the vast majority of known exoplanets via ground and space-based surveys like e.g., WASP \citep{2006PASP..118.1407P}, KELT \citep{Pepper2007}, CoRoT \citep{2009A&A...506..411A}, Kepler/K2 \citep{Borucki2010,Howell2014}, the HATNet project \citep{Bakos2018}, NGTS \citep{2018MNRAS.475.4476W}, and TESS \citep{Ricker2015}.

The wide-field transit surveys conducted thus far have been limited to visible wavelengths ($\lambda < 1\, \mathrm{\mu m}$), an ideal range for the detection of exoplanets orbiting main-sequence stars. 
For example, the ground-based MEarth survey of nearby M dwarf stars \citep{2012AJ....144..145B} uses a custom 715\,nm longpass filter that is sensitive up to $\lambda=1000$\,nm, similar to the Sloan $i+z$ filters. The TRAPPIST survey, another ground-based survey targeting M dwarf stars \citep{2011Msngr.145....2J}, has a peak sensitivity of $98\%$ at about 750\,nm and of $40\%$ at 950\,nm.

To expand the census of transiting exoplanets to hosts of even lower masses, it is necessary to extend transit observations into the near-infrared (NIR) \citep{2019PASP..131k4401T,2021A&A...645A.100S} where low mass hosts are the brightest. 
Surveys in the $\lambda = 1$--$2 \, \mathrm{\mu m}$ range can more easily detect late M dwarfs, BDs and FFPs.
Ground-based IR transit investigations are limited by the bright and variable sky background and are therefore feasible only for the brightest, typically nearby BDs and FFPs.
A few NIR ground-based transit surveys have been recently initiated with the aim of detecting exosatellites orbiting FFPs, BDs, and late M dwarfs \citep[e.g., PINES and SPECULOOS;][]{2019PASP..131k4401T,2021A&A...645A.100S,2022AJ....163..253T}. %Because FFPs are cool
Yet, most nearby BDs and FFPs that can be monitored from ground-based facilities are not clustered in one location, thereby requiring these ground-based transit surveys to conduct time-intensive target-by-target observations. 

Space-based observations, which are sensitive to less luminous and more distant targets, present a unique opportunity to detect transits of moons and satellites orbiting FFPs \citep{Limbach2021} and possibly even tidally heated satellites without transits \citep{2013ApJ...769...98P}. 
Space observations are unaffected by sky background noise and photometric variability due to the Earth's atmosphere, which are the dominant noise sources of ground-based NIR transit searches for target magnitudes greater than $K\gtrsim15$\,mag  \citep{2014ApJ...793...75R,2019MNRAS.483..480V}. 
The Nancy Grace Roman Space Telescope (originally known as WFIRST and hereafter referred to as Roman) is an upcoming space-based, wide-field, NIR telescope\footnote{EUCLID is capable of performing an exosatellite survey; see section \ref{euclidsec}. The Chinese Space Station Telescope (CSST) may also be capable of doing so with an infrared (IR) ($>1\,\mathrm{\mu m}$) imager.} that is ideal for performing an exosatellite transit survey of late M dwarfs, BDs, and FFPs \citep{Spergel2015} due to its unprecedented NIR photometric sensitivity, spatial resolution, field of view and stability. 
Unlike ground-based telescopes and low-Earth orbit observatories, continuous observations uninterrupted by a diurnal cycle are possible with Roman.
The ideal observation field for such an investigation would be a young, densely populated, and nearby star cluster.
Age is an important factor to consider, as FFPs and BDs cool and significantly diminish in brightness over time.
Therefore, only young sources (see Section~\ref{ageLimits} for discussion of detectable ages) are sufficiently bright to permit photometric monitoring at 1--2\,$\mathrm{\mu m}$.

Of all the local, actively star-forming regions, the Orion Nebula is the closest to our home planet at a distance of 400\,pc \citep{2020ApJ...896...79R} that hosts both low- and high-mass star formation. 
While the oldest stars within the Orion Nebula are approximately 12\,Myr, more recent star formation is represented by members of the densely populated Orion Nebula Cluster (ONC), where stars range between 1--3\,Myr in age \citep{Jeffries2007, 2010ApJ...722.1092D}. Extremely young, massive ($\approx$10\,\MJ{}) FFPs have an effective temperature of $\approx 2000$\,K and are bright in the NIR, thus this young population of FFPs are particularly well-suited for observations with Roman.
Therefore, the ONC offers the precise conditions required to facilitate a survey of extrasolar satellites, moons, and planets transiting young FFPs, BDs, and very low-mass stars.

In this paper, we present design considerations of the Transiting Exosatellites, Moons, and Planets in Orion (TEMPO) survey---a proposed 30-day photometric investigation leveraging the Nancy Grace Roman Space Telescope with the aim of detecting a large population of young transiting companions in the ONC. 
If executed, TEMPO would offer the first opportunity to detect a population of exomoons
and exosatellites orbiting BDs and planetary-mass objects, while also detecting a population of small planets transiting young (1--3\,Myr) M dwarfs.
 %(see Figure~\ref{Orion_FOV}).
In this work, we describe the expected detection yield calculations of the TEMPO survey.
An in-depth discussion of the core and ancillary science goals that would accompany the TEMPO survey are described in \cite{Limbach_Soares_InPrep2}.

Section~\ref{design} provides an overview of the TEMPO survey design and a description of the rationale surrounding the design choices. 
Section~\ref{DetectLim} describes a series of simulations to determine the survey's detection limits. 
Section~\ref{EstYields} describes our calculations of TEMPO detection yields. Section~\ref{results} discusses the results of these calculations. 
Section~\ref{Discussion} describes uncertainties in the expected detection yields, as well as the scientific impact of such a survey. 
In Section~\ref{euclidsec}, we briefly discuss and contrast a comparable exosatellite survey with the Euclid Space Telescope. 
Section~\ref{summary} provides some concluding remarks.

%%%%%%%%%%%%%%%%%%%%%%%%%%%%%%%%%%%%%%%%%%%%%%%%%%%%%%%%%%%%%
\section{Survey Design}\label{design}
While the proposed TEMPO survey presents valuable scientific contributions to a wide range of astrophysical subdomains (see the second paper in this series; \citealt{Limbach_Soares_InPrep2}), the survey design optimization is driven by the core science objective: the detection of transiting planets, satellites, and moons in the ONC.
To guide the reader, we summarize the TEMPO survey parameters and expected yields in Table~\ref{tab:para}.

\begin{table}[tbh!]
    \centering   
    \begin{tabular}{c|c}
    \multicolumn{2}{c}{The TEMPO Survey}\\
    \hline
    \multicolumn{2}{c}{Observational Parameters}\\
    \hline
    Field of view & 0.28\,deg$^2$  \\
    Duration & 2 $\times$ 15\,days\\
    Spectral band & F146 ($0.93 - 2.00\,\mathrm{\mu m}$)\\ %or F213 (1.95 - 2.30 $\mathrm{\mu m}$) \\
    Exposure time & 6\,reads, 18\,seconds\\% (continuous observations)\\
    Magnitude limit (1-month) & 29.7\,mag$_{\rm AB}$\\% (50$M_\Earth$ at 1\,Myr)\\
    Photometric precision (1-hr) & 850\,ppm (21\,mag$_{\rm AB}$) \\
    Photometric precision (1-hr) & 125\,ppm (17\,mag$_{\rm AB}$) \\
    \hline
    \multicolumn{2}{c}{Estimated Number of Monitored Sources} \\
    \hline
    Field stars & 5,000\\
    Young stars (ONC) & 2,000 \\
    BDs (ONC) & 560 \\
    FFPs (ONC) & 400 \\
    \hline
    \multicolumn{2}{c}{Estimated Number of  Transit Detections} \\
    \hline
    Young star (Orion) exoplanets & 143 ($100-200$)\\
    Exoplanets (BD host) & 54 ($8-100$)\\
    FFP exomoons/satellites & 14 ($3-22$)\\
    \hline
    \end{tabular}
    \caption{Observational Parameters for the TEMPO survey, which probes the Orion Nebula Cluster (ONC) and neighboring regions. The F146 spectral band was used in the calculation of the magnitude limit and photometric precision. Note that the large range in transit yields of BD and FFP satellites is the result of unknown parameters regarding this young population.}
    \label{tab:para}
\end{table}

\subsection{Target Field Selection}\label{DetectLimb}

There are large populations of relatively bright FFPs and BDs in densely populated, young star-forming regions such as the ONC, Westerlund 2 (2\,Myr) \citep{Zeidler2015}, and the Perseus OB2 association (6\,Myr) \citep{dezeeuw1999}. 
We identify several reasons why the ONC is uniquely well-suited to a transit search with Roman. 
Firstly, it is densely populated --- in the core region, stellar density estimates are as high as $10^4$~stars/pc$^3$; \citealt{1998ApJ...492..540H}. Secondly, it is close enough to permit high photometric precision on even faint FFPs with Roman. In contrast, while young star-forming regions like Westerlund~2 contain more young stars and are more densely populated, source confusion and faintness resulting from their greater distances would hamper the photometric precision required to detect transiting exosatellites around BDs and FFPs.

The ONC is the \textit{only} nearby region that is sufficiently dense, such that we can expect a Roman survey to produce a high exosatellite transit yield. We estimate that Roman would simultaneously monitor approximately 8,000 sources within a 0.28\,deg$^2$ FOV, nearly half of which would be young ONC members. Monitoring such a field permits the detection of dozens of planets, satellites, and moons, while also allowing for the removal of systematic effects via ensemble analysis \citep[e.g.,][]{stumpe}.

 \begin{figure*}[tbh!]
\centering
\includegraphics[width=0.85\textwidth]{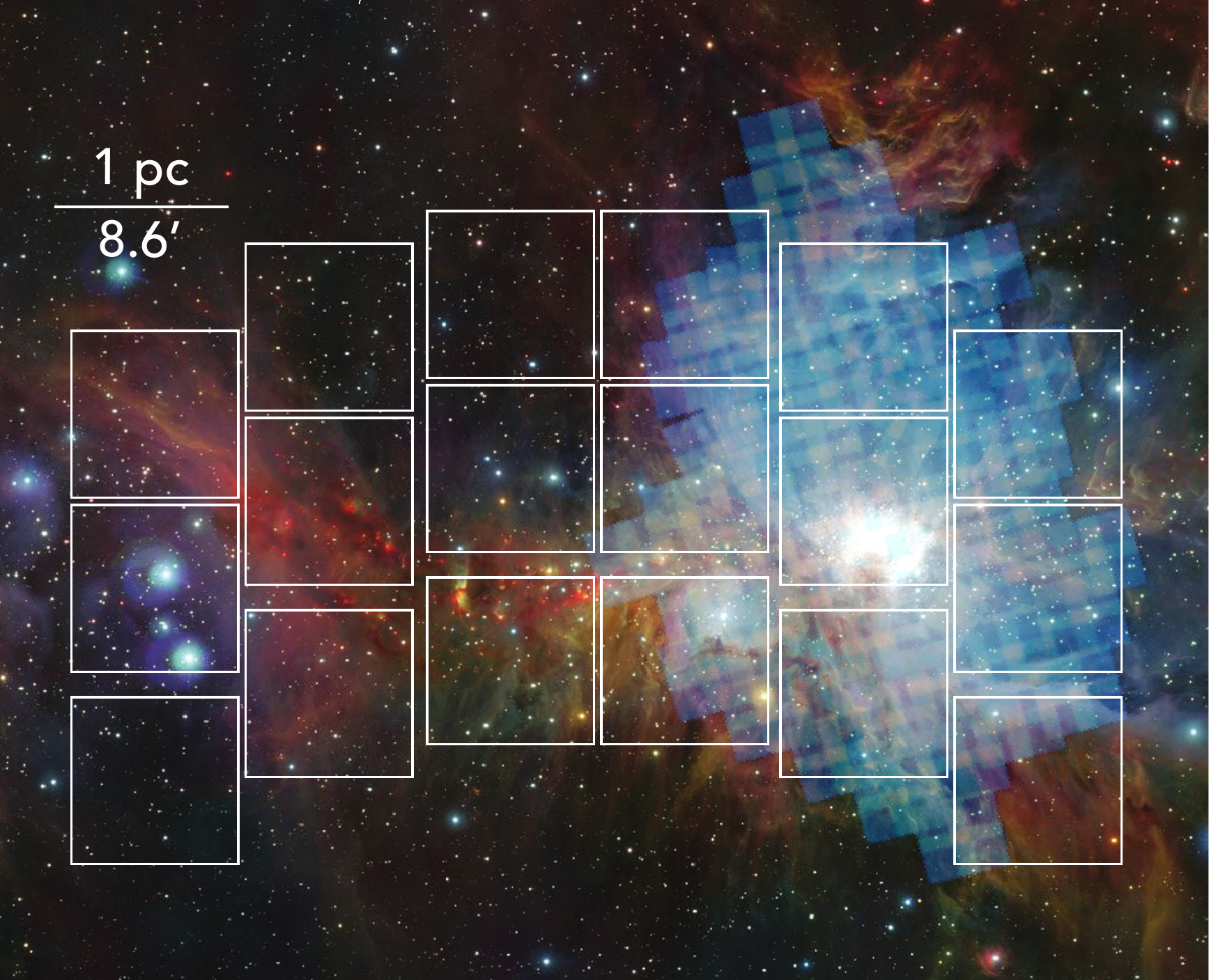}
\caption{The Roman footprint for the proposed TEMPO survey (white outline) and the field coverage from the Cycle~22 Hubble Space Telescope (HST) Treasury Program \textit{``The Orion Nebula Cluster as a Paradigm of Star Formation"} (GO13826, PI: M.~Robberto, 52 orbits) \citep{2020ApJ...896...79R} (blue).
The overlap of the two fields contains the ONC. The background image of Orion was obtained by the Visible and Infrared Survey Telescope for Astronomy (VISTA) infrared telescope. Credit: ESO
}
\label{TEMPOFOV}
\end{figure*}

\subsection{Field of View Orientation Selection}
The ONC is about 7\,pc across, while the surrounding Orion molecular clouds extend 100\,pc across. TEMPO will capture the ONC and a small fraction of the nearby Orion molecular clouds.
We illustrate the TEMPO FOV in Figure~\ref{TEMPOFOV} (white outline).
The TEMPO survey FOV covers a region nearly two times larger than that covered by two HST treasury programs on the ONC \citep{2013ApJS..207...10R,2020ApJ...896...79R}.
Moreover, the proposed TEMPO observation window permits the detection of objects one hundred times fainter (due to the broader band imaging and an extremely long total integration time). 

Although there are no prior wide-field ONC transit surveys with the sensitivity of TEMPO, there have been a number of photometric monitoring campaigns targeting this region. The ONC has been observed by the TESS mission\footnote{The coarse spatial resolution of TESS, 21 arcsec/pixel compared to Roman's 0.11 arcsec/pixel, inhibits transiting companion detection in the dense ONC field.}, and there have been ground-based surveys for planets and eclipsing BD systems. 
Examples include \cite{2006Natur.440..311S} and \citet{ptfo}. 
There have also been several studies that have constrained the number of FFPs, BDs, and stars in the Orion Nebula. This includes investigations with VLT/HAWK-I \citep{2016MNRAS.461.1734D} and HST \citep{2020ApJ...896...79R,2020ApJ...896...80G}. \cite{2020ApJ...896...80G}, in particular, achieved a faintness limit of $21$\,mag$_\mathrm{{AB}}$ and covered a 13\,arcmin$^2$ FOV, detected 1,200 young stars ($< 1.4$\,\Msun{}), 320 BDs, and 220 FFPs. TEMPO would be capable of monitoring all such objects for transits, as well as many fainter, undetected sources contained in the FOV. 

The central coordinates (galactic coordinates 208.821$^{\circ}$, -19.261$^{\circ}$) and orientation of the TEMPO FOV were selected to maximize the number of monitored FFPs, BDs, and stars.
Leveraging the robust Gaia EDR3 catalog \citep{GaiaDR3}, we calculated the stellar densities associated with varying FOV orientations and central coordinates (all of which contained the core of the ONC). 
We used the Roman FOV footprint provided by the Mikulski Archive for Space Telescopes \citep{MAST}.
To identify likely ONC members, our stellar density estimates incorporated Gaia EDR3 sources with G~$<21$\,mag, and included parallax cuts ($2.4 \pm 0.2$\,mas). 

To estimate the number of substellar sources that would be monitored by TEMPO, which are fainter than the Gaia EDR3 limiting magnitude, we scaled the number of substellar sources detected in a prior HST survey \citep{2020ApJ...896...80G}, assuming the same heterogeneous distribution scaled by measured stellar density in TEMPO's larger FOV.
Specifically, we employed the Cycle~22 HST Treasury Program \textit{``The Orion Nebula Cluster as a Paradigm of Star Formation"} (GO13826, PI: M.~Robberto; \citealt{2020ApJ...896...79R}).
The field coverage for this treasury program carried out in the near-IR is shown in Figure~\ref{TEMPOFOV} (blue outlines). 
We employed identical magnitude and parallax constraints in this stellar density calculation.
We find that the TEMPO survey would monitor approximately 2,000 young stars, 560 BDs, and 400 FFPs (see Table~\ref{tab:para}).
This amounts to approximately $1.75$ times the number of cluster members classified by \cite{2020ApJ...896...80G}. 
Additionally, we estimate that TEMPO would monitor approximately 5,000 field stars for transits, as compared to the 2,500 field stars fainter than $m_{130} = 14$\,mag (Vega mag) detected in \cite{2020ApJ...896...79R}.

\subsection{Filter Selection}
\begin{figure*}[tbh!]
\centering
\includegraphics[width=0.9\textwidth]{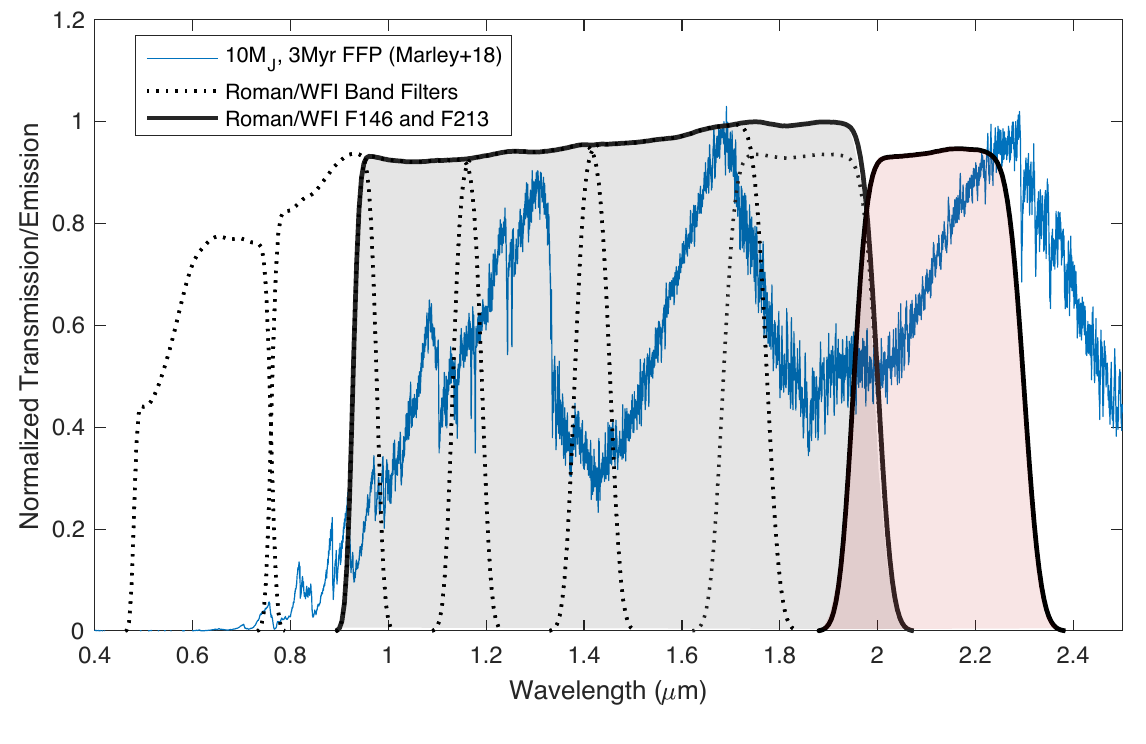}
\caption{The normalized transmission of the Roman broadband F146 filter (black line + gray shaded region), selected for use in the TEMPO survey.
The F213 filter (black line + red shaded region) was also considered in the TEMPO survey design. The F213 filter was found to be suboptimal in detecting the smallest-radius exosatellites.
The normalized emission from a 3\,Myr FFP ($T_{\mathrm{eff}}=2100$\,K, $R=2$\,$\mathrm{R_{J}}$, and $M=10$\,\MJ{}) is shown (blue). 
This FFP emission model is provided in \cite{marley2018}.}
\label{RomanFilters}
\end{figure*}

To maximize sensitivity, TEMPO adopts the Wide-Field Instrument (WFI) with the broad NIR F146 filter ($0.93$--$2.0\,\mathrm{\mu m}$). 
We also investigated the detection yields associated with employing the narrower NIR F213 filter ($1.95$--$2.30\,\mathrm{\mu m}$), finding that this filter is unable to detect the smallest-radius exosatellites that fall within the F146 detection limits. 
Section~\ref{DetectLim} discusses our investigation of the F213 filter in more detail.
Figure~\ref{RomanFilters} shows the broad coverage of the F146 filter (gray-shaded region), as well as the narrower wavelength coverage of the F213 filter (pink-shaded region). 
The F146 filter overlaps significantly with the normalized emission from a 3\,Myr FFP with a mass of 10\,\MJ{} (solid blue line).  
There is little emission from young BDs and FFPs at $\lambda \leq 1\,\mathrm{\mu m}$, thereby preventing optical transit surveys from efficiently monitoring these objects. 

A magnitude range of 17--23\,\ab{} roughly corresponds to 1--5\,Myr sources at a 400\,pc distance with mass ranging between 1\,\MJ{} and 0.2\,\Msun{}.\footnote{We assume extinction is $A_{\rm F146}$=2\,\ab{}.} The majority of FFPs and BDs targets in the ONC fall within this range. High-precision photometry obtained with the TEMPO survey will produce near photon-noise-limited light curves for those targets.

\subsection{Observation Window Selection} 
We devised the TEMPO survey in response to the 2021 \textit{Roman Request for Information\footnote{\url{https://roman.gsfc.nasa.gov/science/Early-definition_Astrophysics_Survey_Option.html}}} (RfI), which limited proposed surveys to an observation window of 700\,hr.
We propose to observe a single FOV, containing the ONC, for the full 700\,hr duration. 
Since the FFP and BD targets in our FOV are generally too faint for Gaia observations, establishing cluster membership using astrometric measurements for these faint targets is critical.
Therefore, we propose that the 700\,hr (29\,d) observation window be split into two 350\,hr (14.6\,d) sub-observations separated by one year.
This is an important design consideration, as establishing cluster membership is critical to the core science goals. 
We refer the reader to \cite{Limbach_Soares_InPrep2} for more discussion on the core and ancillary science outcomes of the TEMPO survey.

We propose continuous monitoring of the field (without slewing) during both 350\,hr periods, taking continuous 18\,s exposures throughout both cycles.
It is possible to detect multiple transits of FFP/BD satellites with orbital periods of several days during the 14.6\,d sub-observation windows. 
Detecting multiple transits is essential for disentangling the effects of host variability from transit signals, which has been shown to be a critical challenge for transit detection \citep{2017AJ....154..224R,2017AJ....153...64M,2018AJ....155....4M,2020AJ....160...33R,Limbach2021,2021AJ....161...23M}. 

In cases where a single transit is detected in each 15\,d set of observations, methodologies developed from K2 and TESS investigations will help to constrain orbital parameters \citep{becker2019, Dholakia} and enable efficient transit confirmation with ground-based telescopes or the James Webb Space Telescope (hereafter JWST) \citep{Gardner2006}. 
Besides astrometry, the significant temporal gap in sub-observations will also greatly assist in the vetting of transit candidates, as transit signatures are strictly periodic, while host variability is observed to evolve with time, which should be apparent with comparisons between sub-observation windows that are one year apart. The effects and mitigation of host variability are discussed more in detail in Section~\ref{HostVar}.

\subsection{Exposure Time and Cadence Selection}
To minimize detector saturation concerns, we use 18\,s exposures, which is the shortest possible exposure time for precision photometry with Roman.
This short integration time would permit the monitoring of sources as bright as 17\,\ab{} with the F146 filter, while achieving near-photon limited monitoring of sources as faint as 23\,\ab{}.
In the case of the F213 filter, the 18-second exposure would correspond to a brightness limit of 15\,\ab{}.
By co-adding the full 15\,d observations, it will be possible to detect sources as faint as 30\,\ab{} at a statistical significance of 5$\sigma$.
This corresponds to the direct-imaging detection of candidates that are $1.0$\,Myr FFPs with sub-Saturn masses down to $50$\,\ME{} \citep{2019A&A...623A..85L}. 
Additional confirmation steps would be needed to confirm these candidates, as field interlopers will vastly outnumber real cluster members and can produce false positives.
This is a particularly important consideration given the small proper motion values associated with ONC sources.

\section{Simulations to Determine Exosatellite Detection Limit}\label{DetectLim}

In this section, we describe the methodology used to determine the detectability of exosatellites transiting FFPs, BDs, and stars (hosts ranging between 2\,\MJ{} and 1.0\,\Msun{}) in the ONC.
Our goal is to investigate transit yield constraints with the F213 \textit{and} F146 Roman filters, in order to determine which of the two filters would produce higher exosatellite detection yields.

\subsection{Transit Detection}
We calculate the F213 and F146 AB magnitude of 2\,\MJ{}-1.0\,\Msun{} hosts by integrating the flux in the F213 (K-band; 1.95--2.30~$\mathrm{\mu m}$) and F146 (broadband; 0.927--2.00~$\mathrm{\mu m}$) Roman spectral bands using theoretical pre-main sequence spectral energy distribution (SED) models for FFPs, BDs, and stars \citep{2008ApJ...689.1327S,Baraffe2015,2016ApJS..222....8D,2016ApJ...823..102C,2018ApJS..234...34P,2021ApJ...920...85M}. 
The F213 and F146 filter throughput are based on the Roman WFI optical design as of October 2021.\footnote{\url{roman.gsfc.nasa.gov/science/WFI\_technical.html}} 
When calculating magnitudes, we assume a distance of 400\,pc to the ONC \citep{2020ApJ...896...79R} and an age of 1.0, 3.0 or 10\,Myr \citep{Jeffries2007}. 
We then employ the predicted Roman WFI F146 or F213 band signal-to-noise ratios (S/N)\footnote{\url{roman.gsfc.nasa.gov/science/apttables2021/table-signaltonoise.html}} to determine the S/N corresponding to the magnitude of each FFP/BD/star after one hour of continuous observing. 

\begin{table}[htp!]
 \footnotesize
\centering
    \begin{tabular}{ccc|cc|cc}
    \multicolumn{3}{c}{} & \multicolumn{2}{c}{K-band} & \multicolumn{2}{c}{Broadband}\\
 \multicolumn{2}{c}{Mass} & Radius & F213 & S/N & F146 & S/N\\
 (\Msun{}) & (\MJ{}) & (\Rsun{}) & (\ab{}) & (1-hr) & (\ab{}) & (1-hr)\\
    \hline
0.002&2.1&0.150&23.66&36.7&23.76&204\\
0.003&3.1&0.155&22.58&95.4&22.71&434\\
0.004&4.2&0.161&21.68&205&21.99&681\\
0.005&5.2&0.167&20.99&351&21.44&930\\
0.006&6.3&0.173&20.51&499&21.00&1176\\
0.007&7.3&0.179&20.14&641&20.62&1431\\
0.008&8.4&0.185&19.84&779&20.29&1690\\
0.009&9.4&0.190&19.60&905&20.01&1934\\
0.010&10.5&0.196&19.38&1034&19.76&2189\\
0.011&11.5&0.202&19.20&1151&19.54&2429\\
0.012&12.6&0.208&19.04&1264&19.35&2659\\
0.013&13.6&0.214&18.89&1373&19.19&2880\\
\hline
0.014&14.7&0.219&18.76&1476&19.03&3097\\
0.015&15.7&0.225&18.63&1578&18.89&3310\\
0.016&16.8&0.232&18.52&1680&18.76&3522\\
0.02&21&0.267&18.04&2153&18.24&4498\\
0.03&31&0.373&17.11&3411&17.25&7098\\
0.04&42&0.47&16.51&4610&16.63&8539\\
0.05&52&0.56&16.08&5516&16.19&8676\\
0.06&63&0.63&15.76&6481&15.86&8638\\
0.07&73&0.70&15.51&7283&15.60&8542\\
0.072&75&0.71&15.46&7458&15.55&8482\\
\hline
0.075&79&0.73&15.54&7196&15.54&8471\\
0.08&84&0.76&15.37&7746&15.37&8192\\
0.09&94&0.79&15.24&8196&15.24&7864\\
0.10&105&0.67&15.51&7313&15.42&8284\\
0.11&115&0.69&15.41&7641&15.32&8072\\
0.13&135&0.73&15.21&8301&15.15&7592\\
0.15&156&0.78&15.00&8968&14.99&7044\\
0.17&178&0.82&14.84&9631&14.80& - \\
0.20&210&0.88&14.61&10048&14.63& - \\
0.25&265&0.98&14.29&10344&14.07& - \\
0.30&310&1.05&14.07&10359&13.66& - \\
0.4&428&1.19&13.62&10621&13.34& - \\
0.5&522&1.27&13.36&10399&13.09& - \\
0.6&628&1.36&13.12&9907&12.86& - \\
0.7&736&1.44&12.92&9437&12.66& - \\
0.8&832&1.49&12.78&9106&12.49& - \\
0.9&944&1.56&12.63&8566&12.33& - \\
0.95&994&1.59&12.56&8292&12.18& - \\
1.0&1029&1.61&12.52& - &12.04& - 
    \end{tabular}
    \caption{Host mass (in \Msun{} and \MJ{}) with the corresponding host radii at 3\,Myr. Source magnitudes are given using the F213 and F146 Roman filters, as well as the accompanying S/N for target observations (distance of 400\,pc is assumed).
    Magnitudes are provided without the inclusion of dust extinction.
    Horizontal lines divide host classes into the following groups: FFPs (top), BDs (center), and stars (bottom).}
    \label{tab:mags}
\end{table}

In Table~\ref{tab:mags}, we present the resulting host AB magnitudes and S/N ratios for both bands. 
The S/N of the FFPs and low-mass BDs is significantly higher in the F146 band, however, S/N is comparable in both bands for more massive stars. 
The F146 saturates at a lower stellar mass, but note that this calculation assumed no extinction, so in practice, more massive stars (with non-zero extinction) can be monitored in the F146 band.
Using the calculated S/N, we then determine the minimum detectable exosatellite radius using a conservative statistical significance of $7\sigma$. Throughout this paper, we use this process to calculate the detection limits over a range of satellite and exoplanet radii, orbital periods, transit durations, and a variety of other parameters. 

To calculate the smallest-detectable satellite radius, we assume the companion is an Earth-density satellite orbiting a host at 1.5\,Roche radii, corresponding to an orbital period of 8\,hr \citep{2013ApJ...773L..15R}. This orbital period is motivated by exoplanets \citep{2013ApJ...774...54S,2013AA...555A..58O,2018MNRAS.474.5523S} found on comparable or shorter orbits.
The transit durations are calculated for edge-on satellites using a simple box approximation for the transit. 
We employed SED models and isochrones from the following: \cite{2008ApJ...689.1327S,Baraffe2015,2016ApJS..222....8D,2016ApJ...823..102C,2018ApJS..234...34P,2021ApJ...920...85M}. 

\begin{figure*}[tbh!]
\centering
\includegraphics[width=0.75\textwidth]{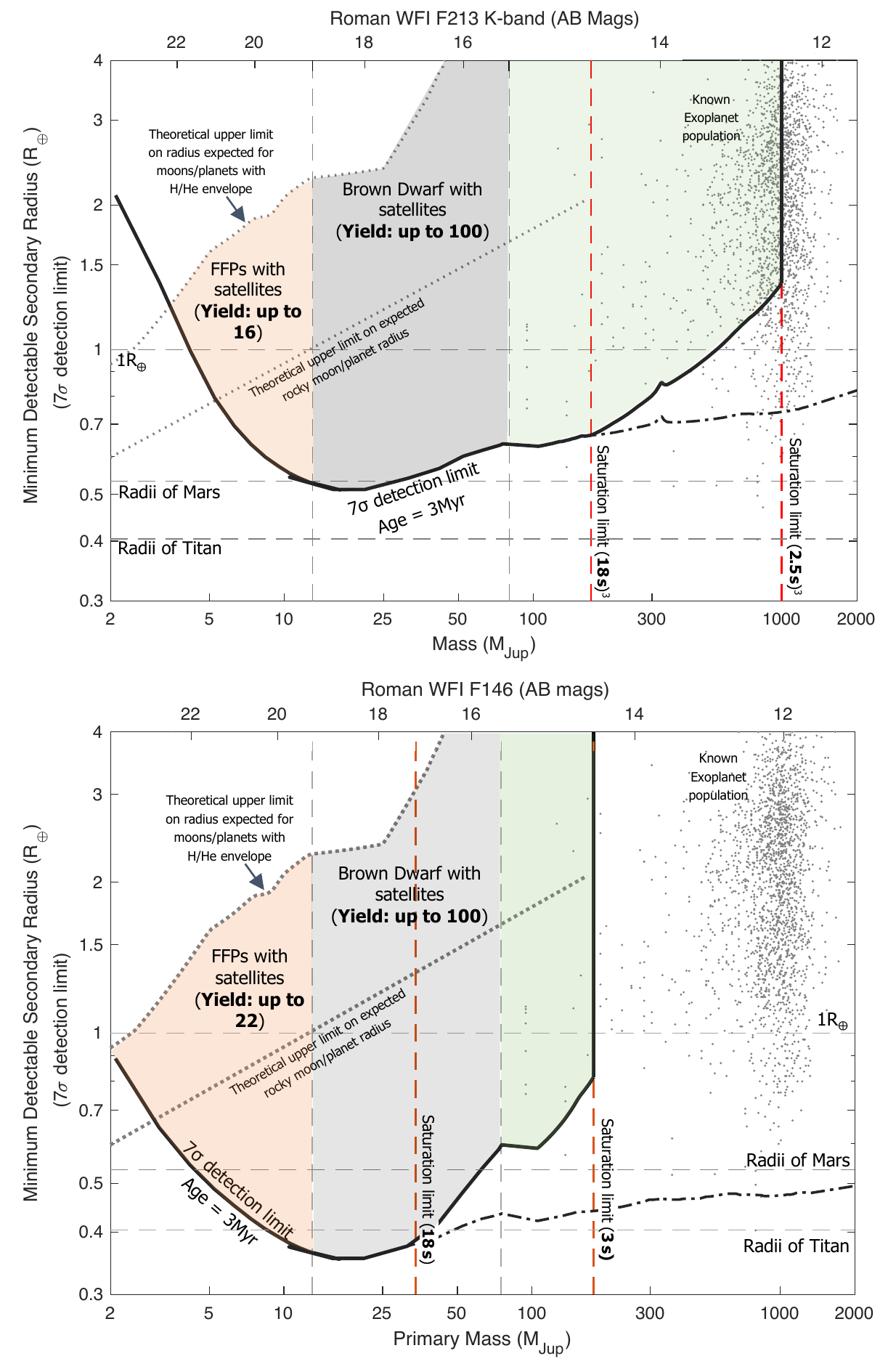}
\caption{{\bf Top:}~Minimum detectable (7$\sigma$) radius (in Earth radii) of an Earth-density satellite with an 8-hr orbital period (y-axis) versus host mass (x-axis, bottom) or AB mag in the Roman/WFI F213 filter (x-axis, top). 
 Detection limits were calculated for host SEDs at 3\,Myr.
The dotted lines depict the expected exosatellite companion theoretical upper limit for (a) rocky satellites and (b) satellites with volatile envelopes (see Section~\ref{Envelopes}). The gray dots indicate the population of known exoplanets.
We included the detection limits in the absence of detector saturation (black dash-dotted line) to demonstrate the sensitivity limit of WFI with defocus. The S/N calculation assumes stacking of the multiple transits observed during the 30\,d survey.
{\bf Bottom:}~Identical to the top panel, however, these calculations utilized the F146 broadband filter.  
}
\label{NoExtinctionWKband}
\end{figure*}

In Figure~\ref{NoExtinctionWKband}, we present the resulting minimum detectable exosatellite radius (solid black line) as a function of host mass (bottom x-axis) and  magnitude (top x-axis). 
The top panel summarizes the results obtained utilizing the F213 filter, whereas
the bottom panel shows the results utilizing the F146 broadband filter.
There is a large parameter space for possible exosatellite detections with the proposed TEMPO survey, as theory predicts exosatellites in the colored regions (peach/gray/green) of Figure~\ref{NoExtinctionWKband} to be common (discussed further in Section~\ref{occurrences}).

Figure~\ref{NoExtinctionWKband} demonstrates that detection down to Mars-sized satellites is possible with the F213 band when companions are on short orbits and minimally extincted. 
In the F146 band, detections down to smaller Titan-sized satellites are possible under these same conditions. 
The ability to detect smaller exosatellites drives the decision to employ the F146 filter for the TEMPO survey.
For reference, Titan is approximately 50\% larger than the Earth's moon and approximately 75\% the size of Mars. 
Transiting short-period satellites the size of Titan, Ganymede, and Callisto with 10--30\,\MJ{} hosts would be detectable in the ONC region assuming minimal extinction.  
The smallest detectable companions ($\approx 0.35$\,\RE{}) would be found transiting 10--30\,\MJ{} hosts. Specifically,
the TEMPO survey would be capable of detecting exosatellites with radii as small as Callisto ($0.38\,R_\Earth$) orbiting a 3\,Myr, 10\,\MJ{} FFP with an approximate effective temperature of 2100\,K and a radius of 2\,\RJ{} \citep{2008ApJ...689.1327S,2021ApJ...920...85M}

\subsection{Detector Saturation Limitations}\label{Saturation}
Our transit detections are limited by a combination of two noise sources: photon noise and host variability (we discuss this latter noise source in more detail in Section~\ref{HostVar}). 
The wide F146 spectral band filter, permitting the collection of as many photons as possible, is effective in reducing the photon noise limit. On the other hand, 
while the F146 filter is more sensitive to smaller transiting satellites than the F213 filter, as shown in Figure~\ref{NoExtinctionWKband}, there are three complicating factors worthy of consideration: the saturation limits of the Roman WFI H4RG detectors; dust extinction, which decreases with wavelength; and nebular emission, which increases in the redder Roman filter. They are discussed in the three following sections.

The Roman Teledyne H4RG-10 IR detectors collect signals via a ``sample-up-the-ramp" technique, whereby each frame is read out non-destructively. To reduce data volume,
multiple frames can be grouped and averaged to reach the desired integration time.
This technique mitigates noise, is effective against cosmic rays, and increases the dynamic range. 
The detectors have a non-adjustable full frame read time of 3\,s. As a result of this operational scheme, each exposure requires at least two frames: one frame for reference and one frame for measurement. 
A minimum of six frames (18\,s integration) is recommended to obtain a precise measure of the count rate.

An 18\,s exposure leads to a saturation limiting brightness of 17\,\ab{} in the F146 filter and 15\,\ab{} in the F213 filter.
For a minimally extincted ONC target, this corresponds to a 40\,\MJ{} BD in the F146 filter and a 0.16\,\Msun{} ONC star in the F213 filter. Therefore, while the F146 filter is better suited for detecting the smallest transiting exosatellites, the saturation limit of this filter will prevent the detection of exosatellites around some high-mass BDs.

Sources between $14.5$--$17\,{\rm mag}$ in the F146 filter and $12.5$--$15\,{\rm mag}$ in the F213 filter would remain unsaturated on a subset of the six frame exposures.
Using these individual frames, it is possible to obtain photometry of brighter sources. In the most extreme case, photometry would still be possible using a single frame, corresponding to F213 = 12.5\,\ab{} or F146 = 14.5\,\ab{} targets.
Beyond these limits, it may still be possible to perform techniques such as halo photometry \citep{White2017} on \textit{all} targets, regardless of their brightness, which may allow for photometric monitoring and precision measurements well beyond the range estimated here. The possibility of detecting the modulation caused by transiting objects in the extremely bright regime is hard to estimate without a detailed characterization of the detector performance. The nonlinearity of the response of IR detectors, in particular, reduces the dynamic range as one approaches saturation.
This may represent another factor affecting the discovery space at the bright end.

A possible strategy for observing stars brighter than the saturation limit would be to introduce a small defocus in the optical path of the Roman WFI.  
This would spread the peak photon counts from the star over a handful of pixels, mitigating saturation limits on the brightest targets.
This technique is similar to the JWST NIRCam shortwave defocus mode \citep{2010SPIE.7731E..0CG}, implemented through the insertion of dedicated defocusing elements. However, defocussing affects all sources in the field and therefore has detrimental consequences on the general sensitivity of a wide field survey.

A different option would be the use of the Roman WFI slitless prism, which would completely eliminate saturation issues on more massive BDs and stars, while possibly providing key information on the atmosphere of the transiting objects. 
It would also greatly improve our ability to differentiate between host variability and satellite transits (discussed further in Section~\ref{HostVar}). 
A principal component analysis (PCA) of the early T-type planetary-mass objects 2M2139 and SIMP0136 HST/WFC3 G141 time-series spectra (similar spectral coverage as Roman/F146) showed
that 99.7\% of host variability can be eliminated using a single principal component
spectrum \citep{2013ApJ...768..121A}. 
Following this path, given that the ONC is densely populated, a fraction of the spectra would be unusable due to photometric blending.
The detrending of light curves from stellar and systematic variability has nevertheless made great progress in the context of exoplanet searches.

Since these techniques preclude reaching the sensitivity needed to detect extremely low-mass objects that are the main targets of the TEMPO survey, we regard them as germane to follow-up studies and their detailed analysis goes beyond the scope of this paper.

\subsection{Dust-Imposed Limitations}\label{Dust}
\begin{figure*}[tb!]
\centering
\includegraphics[width=0.8\textwidth]{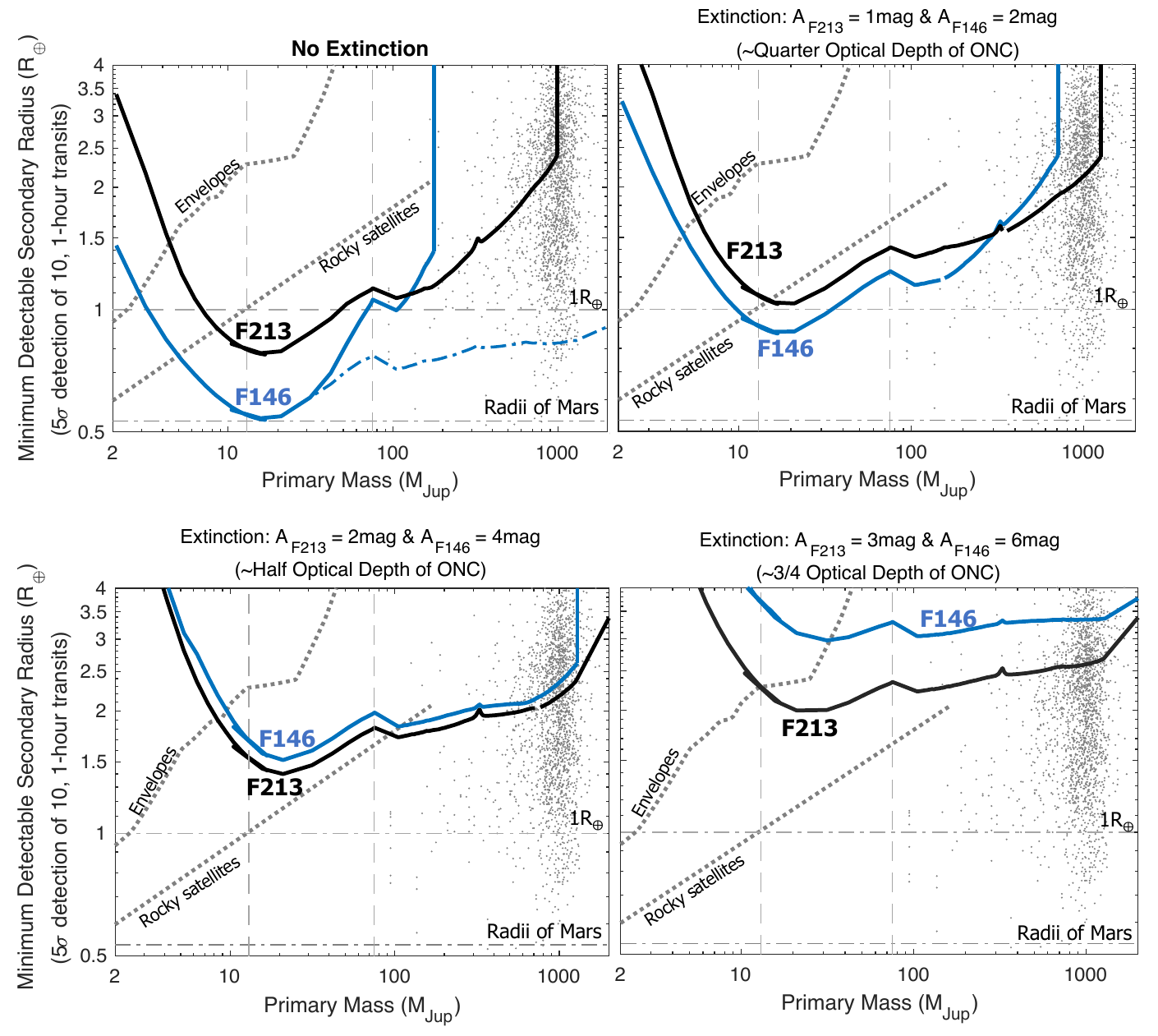}
\caption{Detection limits of transiting exosatellites (3\,d orbital period, edge-on) with the Roman F213 and F146 filters at varying optical depths in the ONC. 
Note that the dust extinction in the F213 band is approximately half that of the F146 filter. 
Therefore, the F213 band outperforms the F146 at larger optical depths. Detection limits on the high-mass end are limited by the saturation of the detector at almost all optical depths in the ONC. In the top left panel, we include the detection limits in the absence of detector saturation (blue dash-dotted line) to demonstrate the sensitivity limit of WFI with defocus. Known exoplanets are shown as small gray dots.}
\label{MinRad2magExt}
\end{figure*}
We use NASA/IPAC Galactic Dust Reddening and Extinction maps\footnote{irsa.ipac.caltech.edu/applications/DUST/} and standard extinction laws \citep{1998ApJ...500..525S,2019ApJ...886..108F} to determine the extinction in our FOV. The dust extinction is almost exactly double in the F146 band than in the F213 band. For our calculations, we assume a uniform extinction across each filter.
The F146 band is more sensitive, but the F213 band has less extinction. This means that the exosatellite detection limits in the two bands become nearly equivalent when $A_{\rm F146}$ = 3\,mag and $A_{\rm F213}$ = 1.5\,mag. 
Given that the extinction in the region covered by our survey is highly non-uniform, the TEMPO survey sensitivity will range widely across the field. The completeness of our search for FFP, BD, and stellar targets will thus depend on both the brightness of the targets and the optical depth throughout the region.

In Figure~\ref{MinRad2magExt}, we illustrate the dust-extincted detection limits for both filters, considering a companion on a 3\,d, edge-on orbit. 
The extinction in our FOV ranges from $A_{\rm F146}$ = 0.2\,mag (near the edges of our FOV) to $A_{\rm F213}$ = 4\,mag and $A_{\rm F146}$ = 8\,mag near the very center of the ONC. The extinction values for the majority of our FOV range between $A_{\rm F146}$ = 0.2-3\,mag.
As shown in Figure~\ref{MinRad2magExt}, severe extinction in the ONC would limit transit detection capabilities of sub-Earth mass satellites orbiting FFPs and BDs to about $A_{\rm F146}$ = 2.7\,mag. In the majority of our FOV, the extinction falls below this threshold, with the exception in the densest regions at the center of the ONC. 

Whereas the F213 filter generally improves the saturation limits and detection limits of sources towards the far end of the ONC, however, the vast majority of FFPs and BDs that are sufficiently bright to monitor for small transiting exosatellites are situated on the near side of the ONC where dust extinction is low (see Figure \ref{Dust}). 
In summary, while the lower extinction in the F213 filter improves the observing efficiency for the most extincted FFPs, the gain does \textit{not} translate into an increase in the exosatellite yield.

\subsection{Nebular Emission Effects on Yields}
\begin{figure*}[]
\centering
\includegraphics[width=0.8\textwidth]{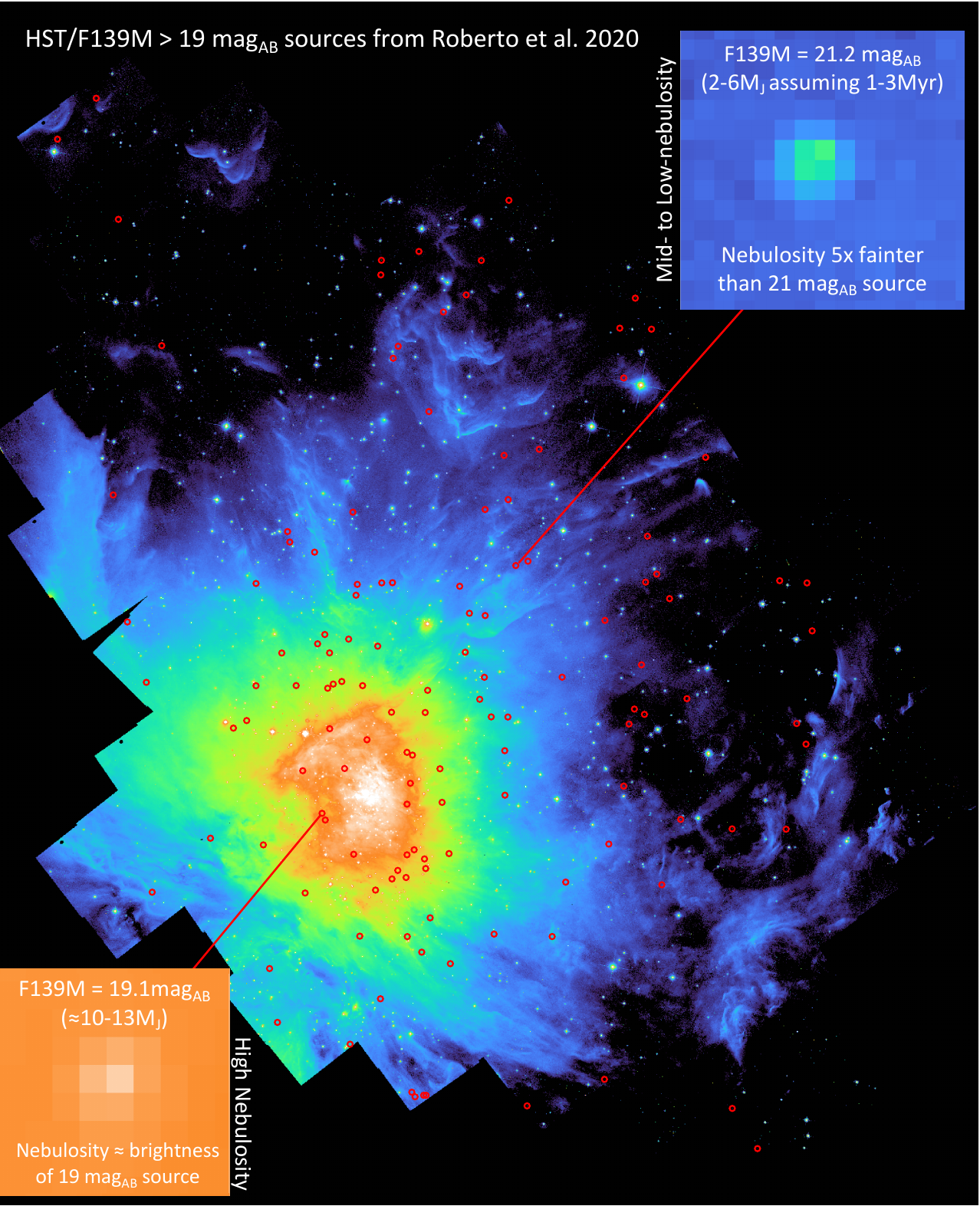}
\caption{Faint (F139M $> 19$ mag$_{AB}$) sources (red circles) associated with the ONC from HST imaging. For comparison, the plate scale of HST/WFC3 and Roman/WFI are 130 versus 110 mas/pixel, respectively, but both instruments oversample the PSF at 1.5$\mu $m. Corner zoom-in panels show the brightness of faint sources relative to nebular emission. In regions of low- to mid-nebulosity (upper right), the luminosity of the nebular emission is significantly less than even the faintest low-mass sources we intend to monitor for transits. In regions of high nebulosity (bottom left), the luminosity of the nebular emission is comparable to a F139M $= 19$\,mag$_{AB}$ source. Data and processing of this imagery is described in \cite{2020ApJ...896...79R}.}
\label{HSTNebulosity}
\end{figure*}

In the highest density regions of the ONC, nebulosity can contribute a non-negligible amount of photon noise to low-luminosity sources. There are two sources of nebulosity: (1) gaseous emission (located at discrete wavelengths), and (2) dust thermal emission (stronger at longer wavelengths). Figure \ref{HSTNebulosity} shows faint (F139M $> 19$ mag$_{AB}$) sources associated with the ONC (data is from the Cycle 22 HST Treasury Program “The Orion Nebula Cluster as
a Paradigm of Star Formation” (GO-13826, P.I. M. Robberto; see \citealt{2020ApJ...896...79R}). 
In regions of low- to mid-nebulosity (see upper right zoom-in), the luminosity of the nebular emission is significantly less than even the faintest low-mass sources TEMPO would monitor for transits. The nebular emission noise contribution is negligible ($<10\%$), and can be ignored. However, in regions of high nebulosity (see bottom left zoom-in), the surface brightness of the nebular emission is $\approx$100\,MJy/str, comparable to the luminosity of a F139M $= 19$ mag$_{AB}$ source (which corresponds to about 10-13\MJ{} for 1-3\,Myr).

\begin{figure}[]
\centering
\includegraphics[width=0.46\textwidth]{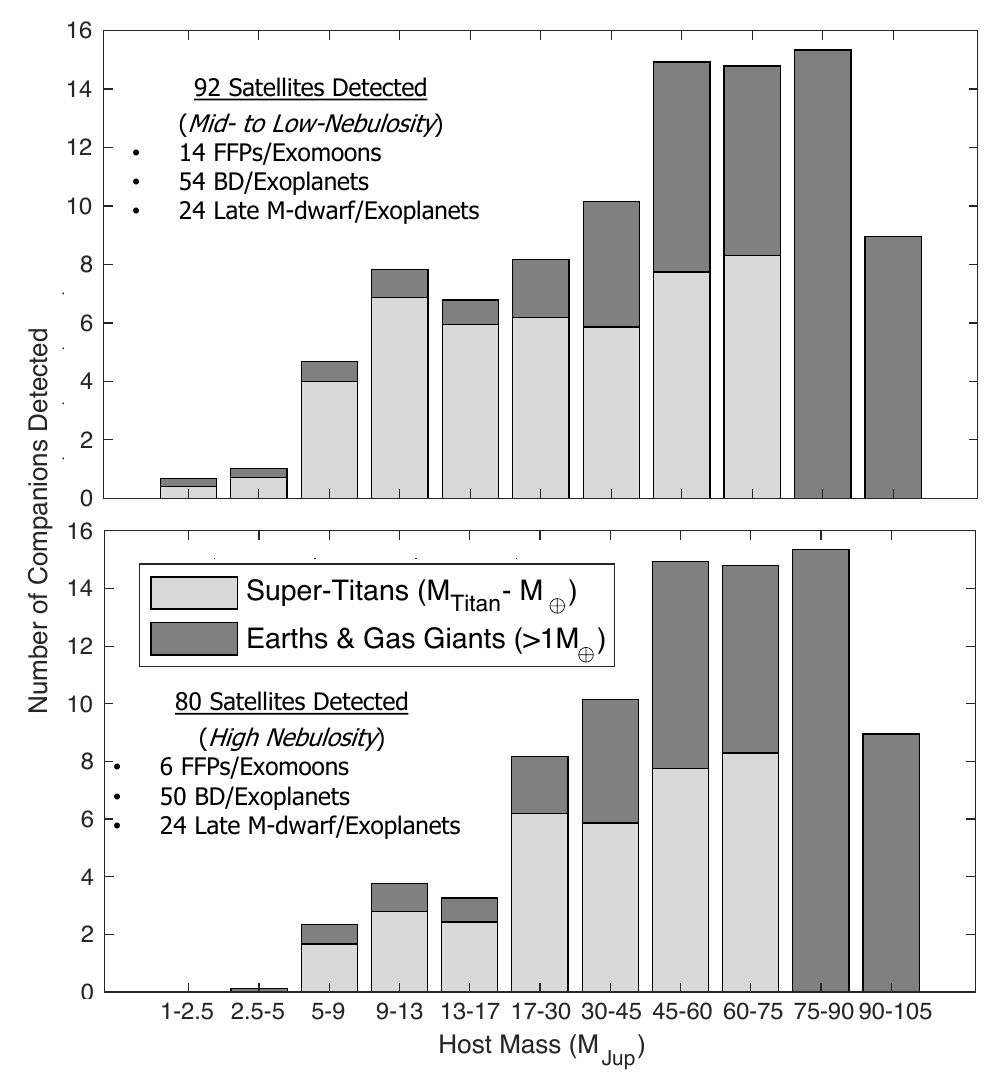}
\caption{Number of expected detections assuming low- to mid-nebulosity (top panel) and a high-nebulosity with surface brightness of 100\,MJy/str (bottom panel) in F146 band using the model parameters described in section \ref{EstYields}.}
\label{NebYields}
\end{figure}

The field of view of the TEMPO survey will contain a relatively low fraction of high-nebulosity regions, implying that in most cases nebular emission will not impact our ability to detect exosatellites. To better understand the emission in high nebulosity regions, we calculate the yields assuming the high nebular emission of 100\,MJy/str. Figure \ref{NebYields} illustrates the expected yields given low- to mid-nebulosity with negligible emission (top) compared to the high emission region (bottom). The FFP/exomoon yields in the high emission region are significantly lower (6 versus 14 exomoon detections), but the BD/exosatellite and stellar/exoplanet detections are unaffected as detectability in the higher-mass regimes is not limited by photon noise.

Although the nebulosity has a minimum impact at F146, it is much more detrimental at F213. At the longer wavelength band, the nebular emission is 2.8$\times$ higher\footnote{Based on the F140M and F210M JWST/NIRCam images from Cycle 1 program “A Census to the Bottom of the IMF in Westerlund 2: Atmospheres, Disks, Accretion, and Demographics”	(GO-2640, P.I. William Best)} and, additionally, the PSF of the source is spread over 2.2$\times$ more pixels. Combined, these effects make it nearly impossible to detect FFP/exomoons in the F213 band in high-density nebulosity regions.

\subsection{Age-Induced Limitations}\label{ageLimits}
In this section, we explore the limitations in exosatellite yields imposed by the age of the system. 
At young ages, the remaining heat from formation makes FFP and BD hosts bright, enabling the transit detection of exosatellites. 
If these hosts are too young, however, their radii can be quite large.
Larger host radii result in smaller transit depths, thereby decreasing the detectability of young companions. 
The smallest exosatellites can be detected when the host is still sufficiently bright from the heat of formation to produce a large S/N, while also having undergone appreciable contraction. Such a host permits exosatellites to produce large, detectable transit depths. 

\begin{figure}[tb]
\centering
\includegraphics[width=0.5\textwidth]{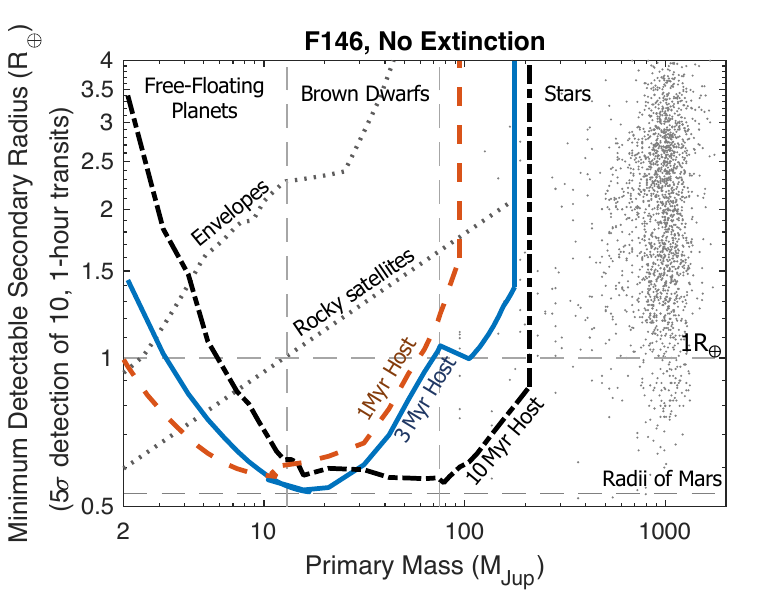}
\caption{Detection limits of transiting satellites with the Roman WFI/F146 filter at three different host ages: 1.0\,Myr (red dashed line), 3.0\,Myr (blue line) and 10\,Myr (black dash-dotted line). Light gray points are confirmed exoplanets. 
The vertical lines separate the host into three regimes: planets, BDs, and low-mass stars. 
The detectability of smaller satellites increases at older ages for BDs, while the detectability of smaller satellites decreases for FFPs with time.
This indicates that the smallest detectable exosatellites observed in ONC will be among massive FFPs and low-mass BDs. 
}
\label{Age}
\end{figure}

Given that the age of the Orion molecular clouds spans 1--12\,Myr \citep{2018AJ....156...84K}, and the ONC spans 1--3\,Myr, we employ SED models at ages of 1.0, 3.0, and 10\,Myr to explore the impact of age on exosatellite detectability \citep{2008ApJ...689.1327S,Baraffe2015,2016ApJS..222....8D,2016ApJ...823..102C,2018ApJS..234...34P,2021ApJ...920...85M}. 
In Figure~\ref{Age} we illustrate the exosatellite detection limits for systems of 1.0, 3.0, and 10\,Myr assuming an edge-on, 3\,d orbital period.  For a given spectral bandpass and host mass, we calculate the optimal host age to detect the smallest companions. 
Contraction and cooling occur quickly for FFPs, resulting in an optimal age that is $\lesssim5$\,Myr.  
For BDs, the contraction process occurs more slowly, resulting in an optimal age for transiting exosatellite detection of $\gtrsim10$\,Myr, as the BD radius is slightly smaller at this age. 
The figure shows that the detectability of smaller satellites increases as the BDs get older, while the detectability of smaller satellites decreases as the FFP hosts age.
In summary, among targets in the 1--3\,Myr ONC, TEMPO will be most sensitive to the detection of small satellites orbiting massive FFPs and low-mass BDs. 

\section{Assumptions for Estimating Yields}\label{EstYields}
In the previous section, we determined transit detectability based on the radius of the exosatellites. 
Estimating the expected transit detection yields requires converting from exosatellite radius to mass and connecting these mass limits to satellite formation models. 
In these calculations, we do not account for the presence of satellites formed via collisionally-induced processes, such as the giant impact theorized to have formed the Earth's moon \citep{Canup2001}, nor do we account for the possibility of moon capture during close encounters between massive FFPs and planetary binaries \citep{2018A&A...610A..39H}. 
We begin with a discussion of host and satellite formation timescales, as well as the observational evidence supporting rapid formation scenarios.

\subsection{Exosatellite, Exoplanet, and BD Formation Timescales}
When estimating yields, we do not attempt to account for the percentage of BD and FFP systems that have not yet completed formation nor do we account for disk dispersal, as these processes are poorly constrained. For very young clusters neglecting these effects may result in overly optimistic predictions.

Observational studies at millimeter/submillimeter wavelengths  offer  support for rapid host and satellite formation timescales. 
For example, \cite{Tychoniec2020} provides strong evidence that there is sufficient dust in young circumstellar disks to form planets within the first $0.5$\,Myr of star formation.
Early-stage satellite formation is further supported by observations of circumstellar disk substructure surrounding the protostar IRS~63 ($\lesssim 0.5$\,Myr), indicating that the planet-formation process begins during this very early evolutionary phase.
Similarly, \cite{2020ApJ...902..141S} found the presence of clear substructure among seven systems aged 0.1--1.0\,Myr. Thus, there is ample observational evidence that at the age of the ONC, exosatellites and exoplanets have had sufficient time to form. 

The Trappist-1 system is analogous in host mass to many of the transiting systems that TEMPO would survey. 
\cite{2022NatAs...6...80R} used n-body simulations to investigate the growth timescale of the TRAPPIST-1 planets, determining that the process was complete in $<$3\,Myr.  
\cite{2017AJ....153..188F} found that 75\% of young stars in the ONC region are diskless, which suggests that planet formation is likely to be complete for the majority of the TEMPO survey stellar targets. However, there is some tension regarding host accretion timescales, as it has also been noted that accretion may occur for longer timescales in some BD and M dwarf systems \citep{2016ApJ...832...50B,2020A&A...638L...6E,2020ApJ...890..106S}.

\subsection{Theoretical Predictions of Young Satellite Envelopes}\label{Envelopes}
Based on \citet{Canup2006} and \citet{2015ApJ...806..181H}, if the secondary forms in an accretion disk around a primary host, the total secondary mass (moon, planet, or satellite) scales as ${\approx}2.5\times10^{-4}~M_{\rm primary}$. These two studies make use of both observational evidence from the Solar system moons \citep{Canup2006} and theoretical simulations \citep{2015ApJ...806..181H} to predict secondary mass.
This upper limit on mass can be converted to a radius using the $R\propto M^{0.28}$ relation \citep{2017ApJ...834...17C}, providing an estimate of the upper limit on rocky satellite radii.
This upper limit is shown throughout Figures~\ref{NoExtinctionWKband}-\ref{Age} by the dotted gray line labeled ``rocky satellite."  
We realistically expect there to be some satellite mass distribution.

In our preceding calculation, we assumed the mass-radius relationship for exosatellites is identical to the observed relation for rocky exoplanets and the solar system's moons. 
This is unlikely to be the case in extremely young systems, as many rocky planets are expected to form with H/He envelopes \citep{2020Icar..33913551L,2021SSRv..217....2S,2021arXiv211204663W}. 
Even companions as small as Mars ($0.1$\,\ME{}) are expected to be sufficiently massive to capture H/He envelopes during their formation (\citealt{1979E&PSL..43...22H,2014P&SS...98..106E,2015A&A...576A..87S,2016ApJ...825...86S}). 
Satellites with H/He envelopes transiting young FFPs and BDs are less challenging to detect given their large radii and accompanying transit depths. 
Since the TEMPO survey targets young, low-mass sources, we expect that these sources will possess some fraction of their initial gaseous envelopes \citep{2021MNRAS.503.1526R}, resulting in a significant increase in the radii of satellites and planets. 

\begin{figure}[tbh!]
\centering
\includegraphics[width=0.45\textwidth]{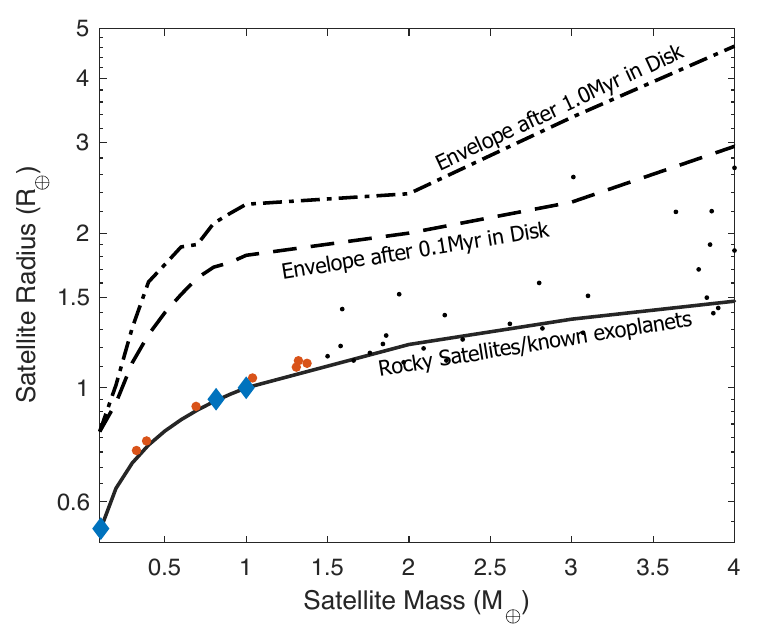}
\caption{Mass-radius relation for satellites $<4$\,\ME{}. 
\textit{Solid black line}: mass-radius relation for rocky satellites from \cite{2017ApJ...834...17C} based on the solar system and known exoplanet population. \textit{Dashed line}: Mass-radius relation expected in the ONC for $<4$\,\ME{} satellites at the time of disk dispersal, assuming satellites are embedded in the disk for 0.1\,Myr.
\textit{Dashed-dotted line}: Mass-radius relation expected in the ONC for $<4$\,\ME{} satellites at the time of disk dispersal, assuming satellites are embedded in the disk for 1.0\,Myr. 
\textit{Red points}: Trappist-1 planets. 
\textit{Blue diamonds}: Earth, Venus, and Mars.
\textit{Black points}: known exoplanets with masses measured to a precision of $\pm30\%$.
Note: all the Trappist-1 exoplanets (and Earth, Venus, and Mars) have lost their H/He envelopes. This is not expected to be the case for ONC exosatellites of similar mass.
TEMPO would allow us to test this theory and our understanding of planet formation.
Exoplanet data were obtained from the NASA Exoplanet Archive (\url{https://exoplanetarchive.ipac.caltech.edu/}).}
\label{EnvelopesFig}
\end{figure}

Theoretical H/He envelopes and mass-radius relations have not been consistently computed in this mass regime at the age of disk dispersal (i.e.,~systems that are only a few Myr old). 
Therefore, we draw upon various computations from the literature to establish a mass-radius relation in this age-mass regime. 
We use the envelope fractions given in \cite{2016ApJ...825...86S} for satellites between 0.1--4\,\ME{}, assuming the satellite is embedded in a disk for either 0.1\,Myr or 1.0\,Myr. 
These envelope fractions and satellite masses from \cite{2016ApJ...825...86S} are converted to satellite radii using the results shown in Figure~3 of \cite{2014ApJ...787..173H}, assuming an envelope entropy of 6.5\,kB per baryon. %($\mathrm{k_B}/B$). 
As an important calibration check, Figure~\ref{EnvelopesFig} indicates that the method used here correctly predicts the Earth's radius was approximately 2\,\RE{} immediately after formation and disk dispersal \citep{1979E&PSL..43...22H,2021MNRAS.503.1526R}.  

The resulting mass-radius relation for 0.1--4\,\ME{} satellites at the time of disk dispersal is shown in Figure~\ref{EnvelopesFig} and incorporated in our transit yield calculations (described further in Section~\ref{results}). In what concerns lower masses, 
our calculations include the  assumptions of \cite{2016ApJ...825...86S} and \cite{2014ApJ...787..173H}, which are aimed at star/planet evolution and may not optimally represent the satellite formation conditions around FFPs and BDs. Our proposed program may provide the 
improved theoretical models and measurements of terrestrial planets, satellites, and moons in their infancy that are required to more accurately predict the mass-radius relation in the 0.1--4\,\ME{} mass range at these early stages.

The radii of young (few Myr) companions more massive than approximately $4$\,\ME{} are likely to be extremely large.
While the exact size will be dependent upon the initial mass, initial entropy, Kelvin-Helmholtz timescale, and boil-off timescale, it is expected to range between that of a gas giant and a stellar radius \citep{2016ApJ...817..107O,2020MNRAS.498.5030O,2021arXiv211009531M}. 
However, we expect most satellites that form around FFPs and BDs to be less massive than 4\,\ME{} based on \citet{Canup2006} and \citet{2015ApJ...806..181H}. 

As these systems evolve, envelope loss is driven by three mechanisms: the high-energy radiation (photoevaporation) from the primary star, core-powered mass-loss \citep{2021arXiv210503443R}, and an initial boil-off stage \citep{2016ApJ...817..107O,2020MNRAS.498.5030O}.
In these theories, the envelopes of less-massive, close-orbiting worlds are often lost over tens to hundreds of Myr, resulting in a more rocky composition at later stages.
This leads to the observed radius valley near 1.7\,\RE{} \citep{Lopez_2014,2017AJ....154..109F,Owen_2017,2020MNRAS.498.5030O,10.1093/mnras/stab895}. 
If the ONC ``proto-terrestrial" planet population is sufficiently young that 0.1--4\,\ME{} mass companions are in possession of H/He envelopes, we would {\it not} expect to detect the existence of the radius valley at approximately $1.7$\,\RE{} in this young population.
The TEMPO survey would allow us to test this directly. 
Such a test would inform our understanding of terrestrial planet formation, offering an opportunity to constrain our theories of Earth's formation, as well as terrestrial exoplanet formation \citep{1979E&PSL..43...22H,Rogers_2011,Mordasini_2012,2016ApJ...825...86S}.

\section{Results}\label{results}
In this section, we present the TEMPO transit detection yields based on several theoretical moon-formation models and measured M-dwarf exoplanet occurrence rates. The range of parameters used to estimate these yields is complete in Table~\ref{YieldSettings}. 
Unless stated otherwise, the plots illustrated in this section employ default parameters, which are provided in Table~\ref{YieldSettings} in bold font.

\begin{table*}[tbh!]
    \centering   
    \caption{Input parameters for yield calculations (default inputs are in bold). {\it References}: [1] \cite{2020ApJ...896...80G}, [2] \cite{2017ApJ...834...17C}, [3] \cite{2016ApJ...825...86S}, [4] \cite{2014ApJ...787..173H}, [5] \cite{cilibrasi2020nbody}, [6] \cite{2019AJ....158...75H}}
    \begin{tabular}{c|c}
    %\multicolumn{2}{c}{Inputs for Yield Calculations}\\
    %\hline
    \textit{Input parameter} & \textit{Allowed values}\\
     \hline
    Filter detection limits & F213$^\mathrm{a}$, {\bf F146}$^\mathrm{a}$ \\ 
    \hline
     WFI defocus$^\mathrm{b}$ &  ON, {\bf OFF}\\  
     \hline
     Host age & {\bf 1.0\,Myr}, 3.0\,Myr or 10\,Myr\\
     \hline
     ONC initial mass function & {$\bf 1.75\times$} known ONC population [1]\\
    \hline
    \multirow{3}{*}{Dust extinction} & Integrated$^\mathrm{c}$, $A_{\rm X}$ (mag) or OFF \\
    & {\bf $\bf A_{\rm F146}$ = 1.2$^\mathrm{d}$ \& $\bf A_{\rm F213}$ = 0.6$^d$} \\
    \hline
    \multirow{3}{*}{Satellite mass-radius relation}&Rocky [2]\\
    & Envelope (0.1\,Myr$^\mathrm{e}$ in disk)  [3,4]\\
     &{\bf Envelope (1.0\,Myr$^e$} in disk) [3,4] \\
     \hline
    Minimum detectable transit  & {\bf Transit depth$ \bf \geq0.05\%$}\\
    due to host variability & Min transit depth range: $0.01-0.1\%$ \\
    \hline
    \multirow{3}{*}{Occurrence rates/satellite mass} & {\bf Theoretical occurrences} [{\bf 5}]\\
    & M-dwarf exoplanet occurrences [6]\\
    & One $5\times10^{-5}M_{\rm Host}$ satellite per host\\
    \hline
    \multirow{3}{*}{Orbital periods} & {\bf Solar system moon periods}\\
    & Theoretical estimates [5]\\
    & Measured mid-M-dwarf distribution [6]
    \end{tabular}
    \\
     \begin{flushleft}
    {{\footnotesize $^\mathrm{a}$ Detection limits are given for a 7$\sigma$ detection of ten, 2-hr satellite transits.}\\
    {\footnotesize $^\mathrm{b}$ Defocusing the WFI would mitigate saturation issues on the H4RG detectors.}\\
    {\footnotesize $^\mathrm{c}$ Yields are calculated at five optical depths through the cloud and averaged.}\\
    {\footnotesize $^\mathrm{d}$ Estimated median $A_{\rm F146}$ of detected satellite.}\\
    {\footnotesize $^\mathrm{e}$ Envelope acquired during time (0.1 or 1.0 Myr) satellite spent in disk during formation.}}
     \end{flushleft}
    \label{YieldSettings}
\end{table*}

The initial mass function (IMF) of the ONC (including the detection of FFPs, BDs, and stars) has been characterized by several previous studies \citep{2016MNRAS.461.1734D,2020ApJ...896...79R,2020ApJ...896...80G}.
For our calculations, we used the IMF mass distribution for FFPs, BDs, and stars in the ONC that is given in \cite{2020ApJ...896...80G}.
We anticipate that the TEMPO survey will detect many more new objects due to the substantially larger FOV and increased sensitivity to survey fainter objects. 
Therefore, for our ``default" yield calculations, we assume $560$ BDs and $400$ FFPs will be monitored for transits. This is 1.75$\times$ the number detected by the HST survey of the ONC performed by \cite{2020ApJ...896...80G}, as described in Section~\ref{design}.  

\subsection{Exomoons and Exosatellites}
\subsubsection{Occurrence Rates and Mass Distributions}\label{occurrences}
The occurrence rate and demographics of exoplanets orbiting stars on short orbits have been well characterized by prior surveys \citep[e.g.,][]{2013ApJ...767...95D, Sanchis-Ojeda2014,Winn2015,Zhu2018,2018AJ....155...89P,2019AJ....158...75H,Vanderspek2019,Uzsoy2021}.
However, the occurrence rate and demographics of the FFP and BD exosatellite population remain unknown, with few constraints measured to date \citep{2018AJ....155...36T, 2021ApJ...922L...2V}. In this section, we focus on estimating the number of exosatellites orbiting FFPs, BDs, and low-mass stars. 

\begin{figure}[t!]
\centering
\includegraphics[width=0.46\textwidth]{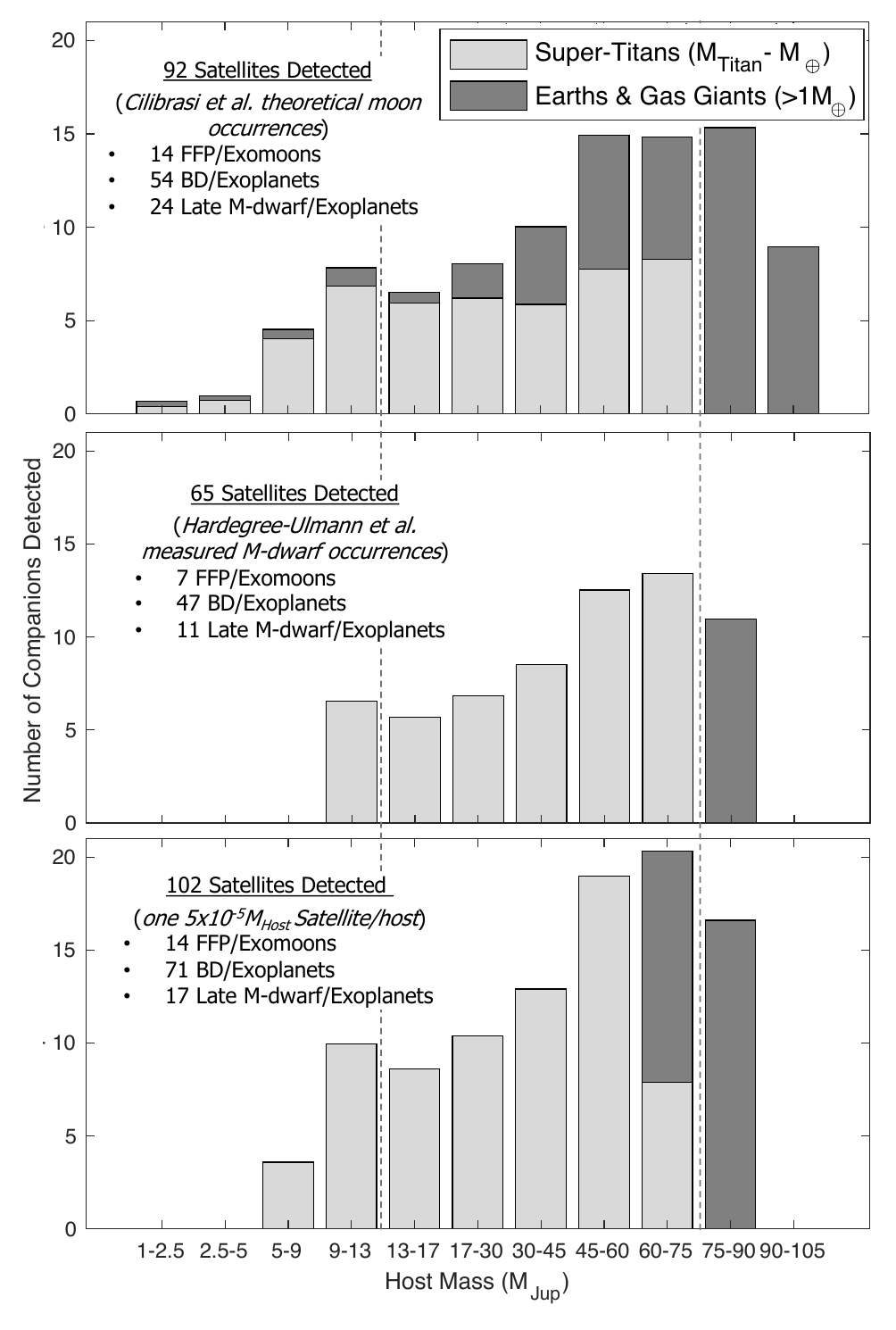}
\caption{The number of expected detections of transiting satellites with the TEMPO survey that accompany three different occurrence rate distributions. \textbf{Top:} using theoretical exomoon formation models \citep{2021MNRAS.504.5455C}. 
\textbf{Middle:} assuming a uniform distribution of one $5\times10^{-5}M_{\rm Host}$ satellite per system.
\textbf{Bottom:} employing the measured mid-M-dwarf exoplanet mass distribution \citep{2019AJ....158...75H}.
Super-Titans ($M_{\mathrm{Titan}}$-\ME{}) are shown in light gray, and satellites more massive than Earth are in dark gray.}
\label{YieldOccur}
\end{figure}

Our estimates are determined using three distinct exosatellite occurrence rates (mass distributions): (a) theoretical exomoon formation models \citep{2021MNRAS.504.5455C}, (b) measured low-mass M-dwarf exoplanet statistics \citep{2019AJ....158...75H}, and (c) a uniform distribution of $5\times10^{-5} M_{\rm Host}$ satellites per system. 
In Figure~\ref{YieldOccur}, we illustrate the detection yields assuming these three different satellite-mass distributions.
Assumptions (a) and (c) result in similar TEMPO survey yields of ${\approx}14$ FFP exosatellites/exomoons and ${\approx}$50-70 BD exosatellites.
Assumption (b)  produces a comparable number of BD companions but roughly half the number of FFP satellites.
The difference may be attributed to the relatively small sample in \cite{2019AJ....158...75H}, with a lack of large exoplanets ($M\gtrsim 5\times10^{-5}M_{\rm Host}$). 
Therefore, the occurrence rates are only valid for the exoplanet mass bins centered on ${\sim}2\times10^{-5}M_{\star}$ and ${\sim}4\times10^{-5}M_{\star}$. 
The \cite{2021MNRAS.504.5455C} models suggest a relatively low, but non-zero occurrence rate for large companions.
Given that these large exosatellites are particularly easy to detect, a nondetection by the TEMPO survey would place strong upper limits on their occurrence rates, thereby constraining the \cite{2021MNRAS.504.5455C} models.

The mass distribution of the satellites varies greatly for the three different occurrence rate models. 
For the \cite{2021MNRAS.504.5455C} theoretical occurrences, we would expect to detect 50\% more super-Titans than $>1$\,\ME{} satellites, whereas for the two other distributions almost all detected satellites will be super-Titans (M$_{\rm Titan}$-\,\ME{}). 
Therefore, the TEMPO survey detection yields would allow us to differentiate between the different occurrence rate distributions. Moreover, TEMPO's observations would help answer open questions such as {\it can Earth-mass exosatellites form around Jupiter and Super-Jupiter hosts?}
It is important to note that if host variability limits the depth of transit detections to $>0.05\%$ ($>$500 ppm), only the largest exosatellites will be detectable around low-mass stars. 
We discuss the impact of host variability in Section~\ref{HostVar}. 

\subsubsection{Expected Orbital Periods}
\begin{figure*}[tb!]
\centering
\includegraphics[width=0.98\textwidth]{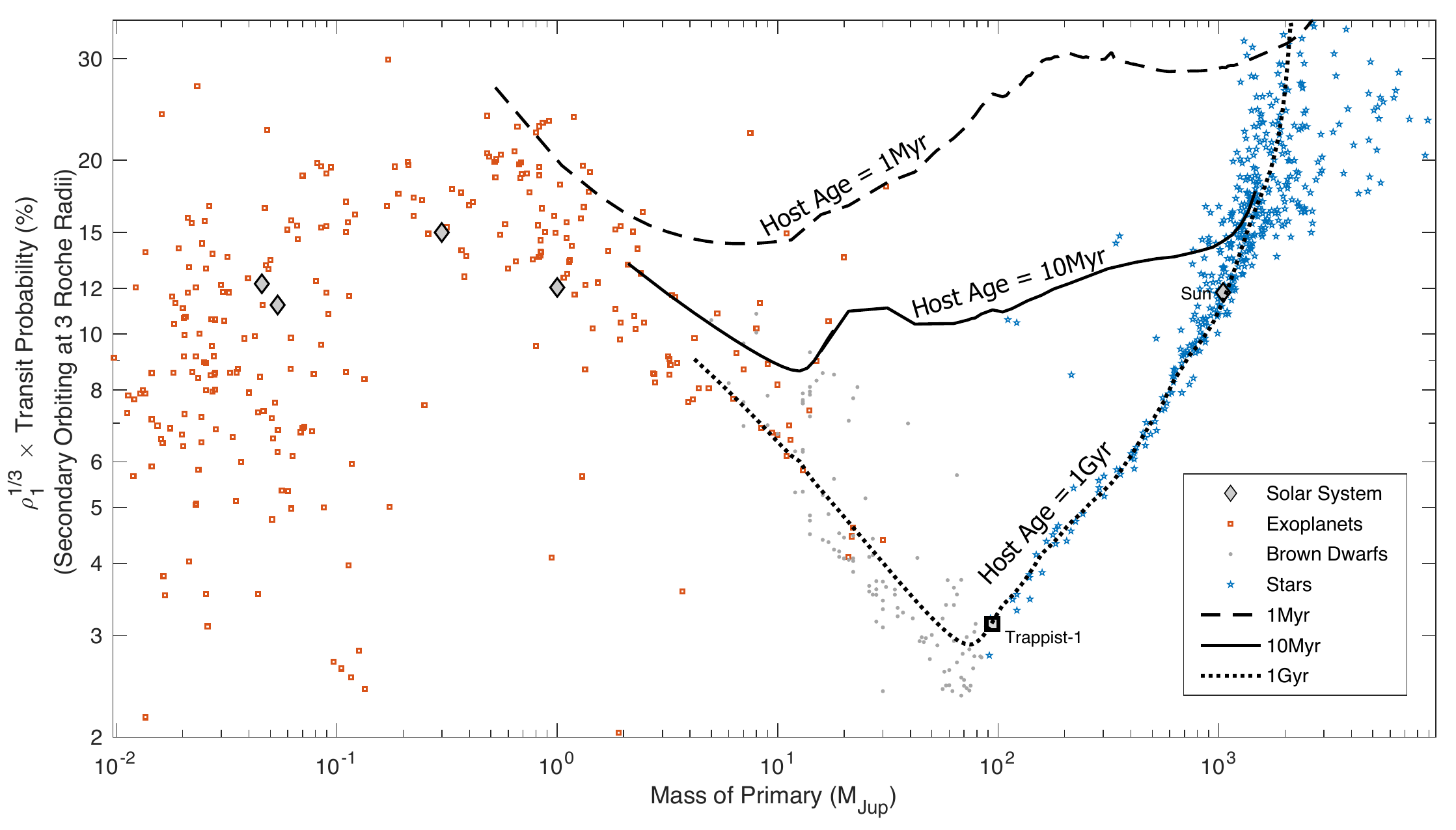}
\caption{Age-dependent geometric transit probability versus primary mass,
assuming a secondary companion orbiting at 3 Roche radii with a density of $\rho_1=1$g\,cm$^{-3}$. The geometric transit probability is shown for systems at three ages: 1.0\,Myr (dashed line), 10\,Myr (solid line) and 1\,Gyr (dotted line).
The transit probability of the companion at 1\,Myr is very high (15-30\%) for all host masses. The transit probability of the companion decreases sharply with age for high-mass BDs and low-mass stars.
The diamonds represent sources in the solar system hosts, including the Sun.
Data were taken from [1] the NASA Exoplanet Archive
(only objects with $a>0.1$AU are included in this plot),
[2] \cite{2009AIPC.1094..924G, best2021} and [3] \cite{Parsons_2018, southworth2014debcat}.}
\label{TransProbWAge}
\end{figure*}

In this section, we incorporate various orbital period distributions and calculate the transit probability accompanying each orbital period in order to predict the fraction of companions that are observable via the transit technique. 
In Figure~\ref{TransProbWAge}, adapted from \cite{Limbach2021}, we illustrate the age-dependent geometric transit probability as a function of primary mass, assuming a secondary companion with density $1$g\,cm$^{-3}$, orbiting at 3\,Roche radii .
The radii of BDs are large at young ages (values listed in Table~\ref{tab:mags}), and this is reflected in the high satellite transit probabilities corresponding to these hosts, with values reaching $30\%$ \citep{2016A&A...588A..34H}. 
For extremely young systems (1.0\,Myr; top dashed black line), the transit probability of companions is high regardless of the host mass. 
At older ages (1~Gyr; bottom dotted black line), there is a local maximum for hosts that are gas giant planets. Transit probabilities decrease sharply with age for BDs and low-mass host stars. 
The original figure by \cite{Limbach2021} shows a the trough in transit probability at the BD-to-low-mass star transition that we do not find at young ages, suggesting again
that young star-forming regions are ideal locations for the detection of transiting companions.
%to these hosts. 

The orbital period distributions used in our analysis include (a) the solar system planet-moon orbital periods (only including solar system moons more massive than $10^{-5}M_{\rm planet}$) (b) the theoretical planet-moon period distribution given in \cite{2021MNRAS.504.5455C}, and (c) the measured star-planet orbital periods of M-dwarf exoplanets from \cite{2019AJ....158...75H}. 
When only orbital separations are given, we compute orbital periods based on the host mass. All these distributions predict, on average, systems containing exosatellites (or multiple exosatellites) on $\lesssim$10\,d orbits. This part of the parameter space, readily accessible with our proposed transit survey, is well separated from the orbital period distribution of the exoplanet population.

\begin{figure}[tbh]
\centering
\includegraphics[width=0.47\textwidth]{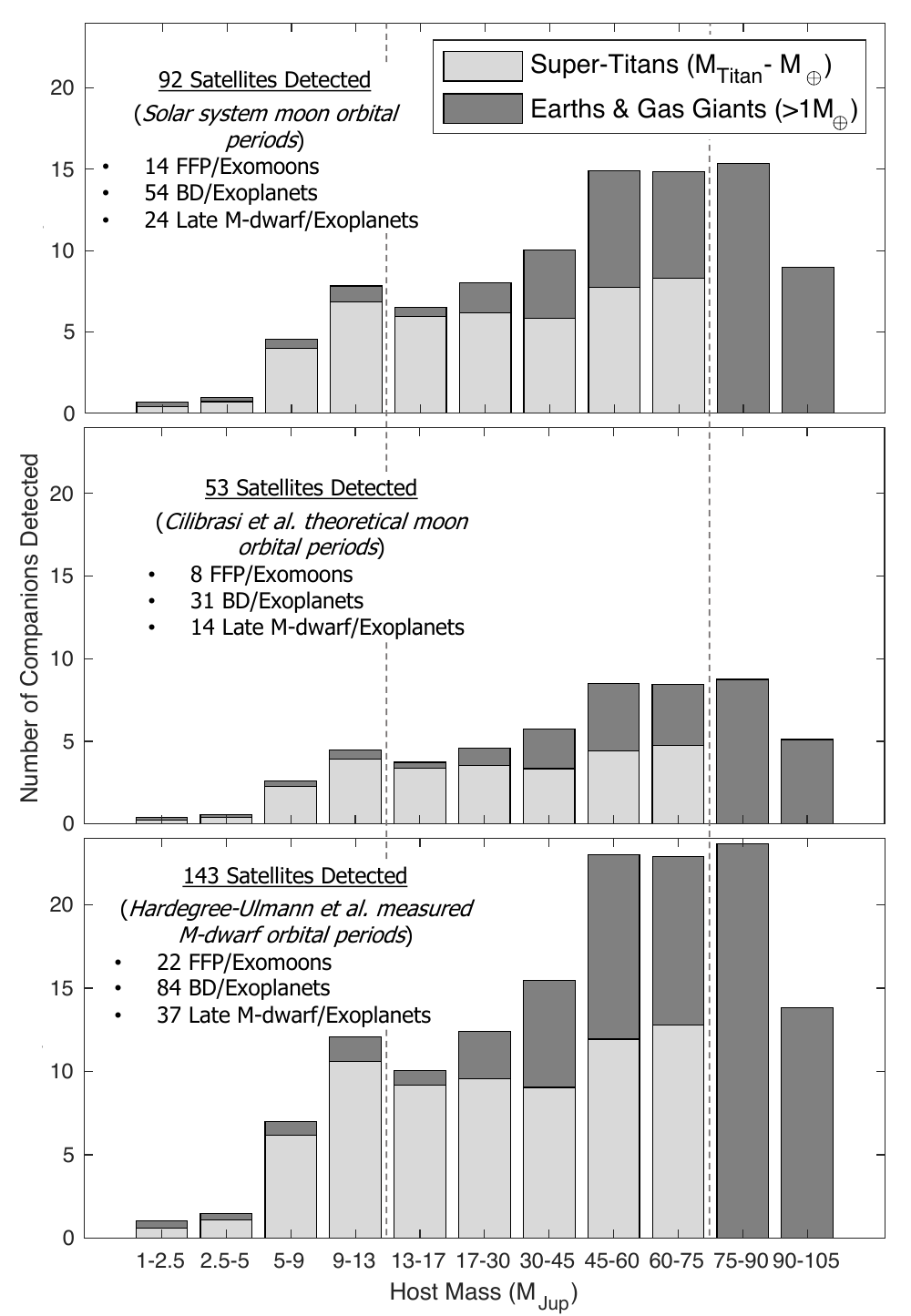}
\caption{Number of expected detections for three distinct satellite orbital period distributions.
\textbf{Top:} Yields using the Solar System planet-moon periods (only including moons more massive than $10^{-5}M_{\rm planet}$).
\textbf{Middle:} Yields using the theoretical planet-moon period distribution given in \cite{2021MNRAS.504.5455C}.
\textbf{Bottom:} Yields using the measured star-planet period of M-dwarf exoplanets from \cite{2019AJ....158...75H}.}
\label{YieldPeriods}
\end{figure}

In Figure~\ref{YieldPeriods}, we illustrate the expected TEMPO yields for the three orbital period distributions.
They produce significantly different yields, summarized in Table~\ref{SummaryOfYields}. 
Using the theoretical orbital period distributions from \cite{2021MNRAS.504.5455C} we predict the detection of only 53 companions, eight of which are expected to orbit FFPs. Conversely, when we use the period distribution from  \cite{2019AJ....158...75H}, we predict the detection of nearly 150 companions, of which 22 would be exosatellites transiting FFPs. 
Finally, the \cite{2021MNRAS.504.5455C} orbital period distribution predicts exomoons orbiting on much longer periods than  observed in the M-dwarf exoplanet population and among solar system moons.
This may be because the exomoons in the \cite{2021MNRAS.504.5455C} simulations did not have adequate time to migrate, whereas the known M-dwarf exoplanets and solar system moons have. 
If the young exosatellites in the ONC have not had sufficient time to migrate, one would expect detection yields similar to those measured using the \cite{2021MNRAS.504.5455C} orbital period distribution. 
Nonetheless, the measured orbital periods of the exosatellites in the ONC with TEMPO would allow us to place limits on the timescales for planet/moon migration and to differentiate between the proposed orbital period distributions. 

\subsubsection{Yields for Rocky Worlds vs. Satellites with Envelopes}
\begin{figure}[]
\centering
\includegraphics[width=0.45\textwidth]{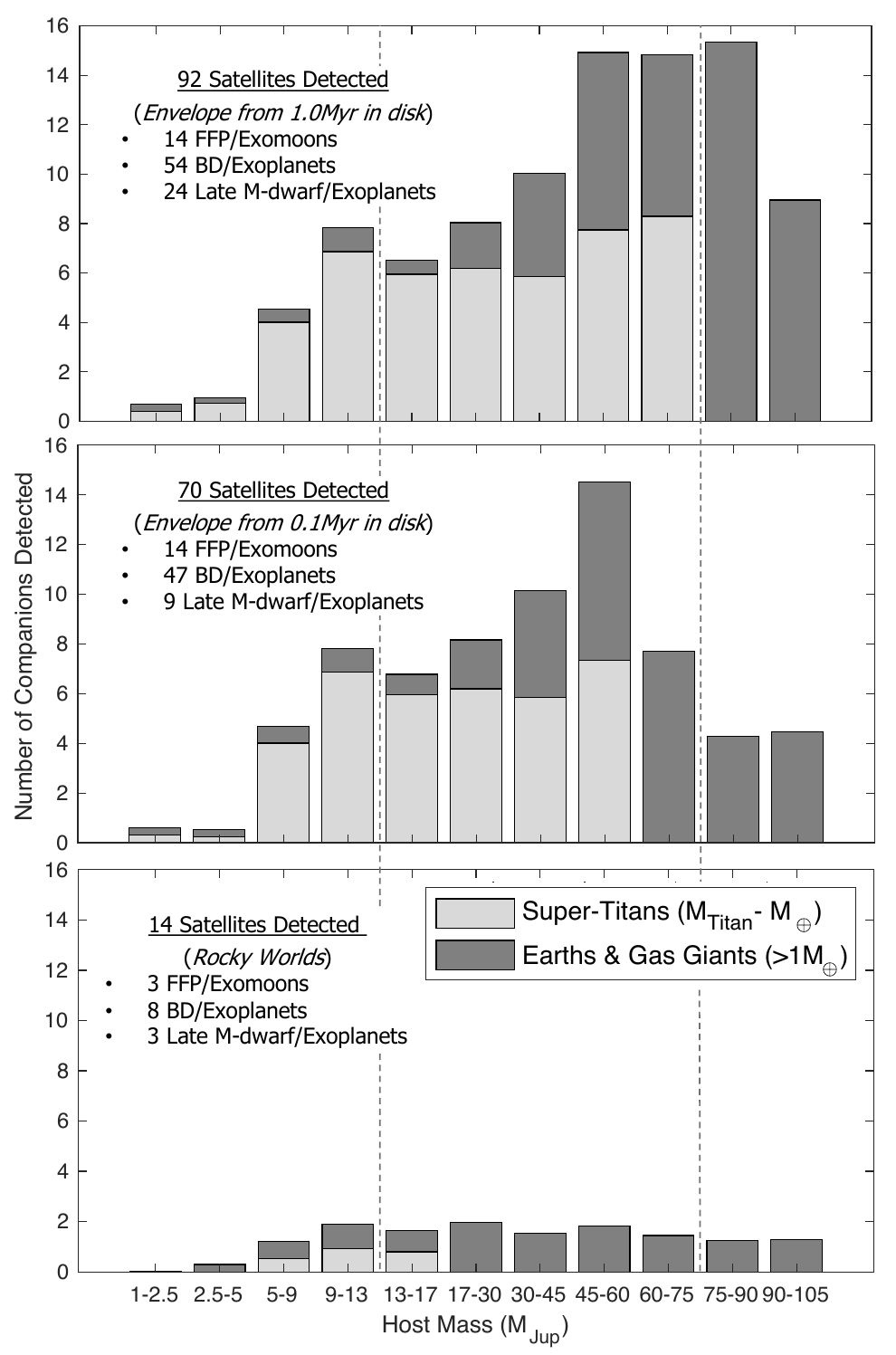}
\caption{
\textbf{Top:} Expected yields for exosatellites that have an envelope captured after being embedded in a disk for $10^6$ years.
\textbf{Middle:} Expected yields for exosatellites that have an envelope captured after being embedded in a disk for $10^5$ years.
\textbf{Bottom:} Expected yields for rocky exosatellites.
}
\label{YieldEnvelopes}
\end{figure}

As discussed in Section~\ref{Envelopes}, young exosatellites with masses $>1~M_{\mathrm{Mars}}$ are expected to possess H/He envelopes at early evolutionary stages. 
However, this assumption is based solely on theory, as there are no known sufficiently young Earth-mass exoplanets (or exomoons) to test it. 
In Figure~\ref{YieldEnvelopes}, we illustrate the expected detection yields for exosatellites that have captured an envelope after having been embedded in a disk for $10^6$\,yr (top),  for $10^5$\,yr (middle), and for rocky exosatellites (bottom panel). 
Figure~\ref{YieldEnvelopes} shows that the TEMPO yield predictions vary widely depending on whether the ONC satellites are in possession of H/He envelopes. 
For the most massive BDs and low mass stars (brighter than F146 $\lesssim 16$\,\ab{}), follow-up transmission spectroscopy with HST or JWST should be possible (modeling of follow-up JWST observations is explored in \citealt{Limbach_Soares_InPrep2}).
These data could be used to investigate the presence of an H/He envelope, offering an opportunity to constrain existing models.

\subsubsection{Host Variability Effects on Yields}
As discussed in Sections~\ref{Saturation} (and later in Section~\ref{HostVar}), we expect host variability and detector saturation to impact the number of detectable exosatellites in our transit search. 
As baseline, we assumed the minimum detectable transit depth due to host variability would be $0.05\%$. In this section, we explore how the detection yield changes with minimum detectable transit depths, exploring depths of $0.01\%$, $0.05\%$, and $0.1\%$. Justification for this range of values is discussed in section \ref{HostVar}.
\begin{figure}[]
\centering
\includegraphics[width=0.46\textwidth]{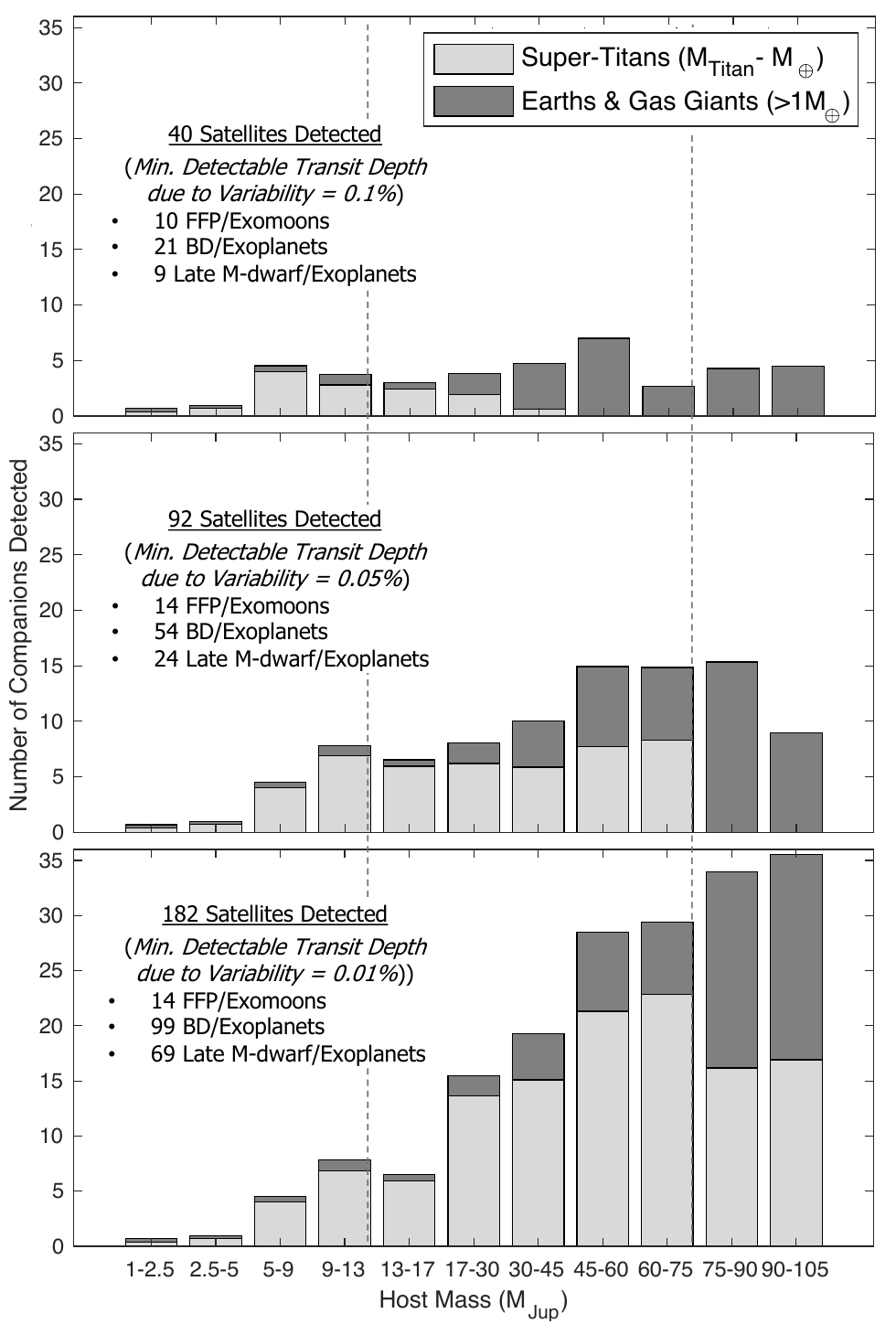}
\caption{Number of expected detections assuming a minimum detectable transit depth due to host variability of $0.1\%$ (top panel), $0.05\%$ (middle panel), and $0.01\%$ (bottom panel). For all calculations, we assume that the WFI is defocused and that there are no detector saturation limits.}
\label{YieldVarDefoc}
\end{figure}
As illustrated in Figure~\ref{YieldVarDefoc}, the number of detected transiting exosatellites significantly increases from 40 (when the minimum detectable transit depth is $0.1\%$) to 182 (when the minimum detectable transit depth is $0.01\%$). 
This implies that the ability to detect smaller transits using state-of-the-art algorithms and synergistic, panchromatic observations with the Vera Rubin Observatory (hereafter Rubin) would significantly improve detection yields.
Moreover, the detection of transits down to $0.01\%$ depths would double the number of detected BD exosatellites detected with the TEMPO survey.

\subsubsection{Summary Table of Expected Yields}\label{summary}
Table~\ref{SummaryOfYields} summarizes the exosatellite yield measurements described throughout the preceding subsections. 
A list of the parameters explored is provided in Table~\ref{YieldSettings}, with default model parameters listed in bold font.

\begin{table*}[]
    \centering   
    \caption{TEMPO exosatellite yield predictions under various assumptions. Yields include companions to FFPs, BDs and late M dwarfs ($<100$~\MJ). The default model parameters are listed in Table~\ref{YieldSettings}, and are F146 filter, defocus off, 1.0\,Myr host age, $A_{\rm F146} = 1.2$\,mag, theoretical occurrence rates of exosatellites based on \cite{cilibrasi2020nbody}, orbital periods based solar system moon statistics, and a minimum detectable transit depth due to host variability of 0.05\%.}
    \begin{tabular}{l|c|c|c|c}
     & \multicolumn{4}{c}{Yield}\\
    Parameter that is varied & FFPs & BDs & Late M dwarfs & Total\\
    \hline
    \multicolumn{2}{l}{\it Min. Detectable Transit due to host variability}\\
    \hspace{2cm} Transit depth $\geq$0.10\% &	10& 21 & 9 & 40\\
    \hspace{2cm} Transit depth $\geq$0.05\% &	14& 54 & 24 &92\\
    \hspace{2cm} Transit depth $\geq$0.01\% &	14 & 99 & 69 & 182\\
    \hline
    \multicolumn{2}{l}{\it Exosatellite envelope}\\	
    \hspace{2cm} 1.0 Myr embedded in disk&	14 &54&24&92\\
    \hspace{2cm} 0.1 Myr embedded in disk&	14&47&9&70\\
    \hspace{2cm} Rocky worlds& 3&8&3&14\\
    \hline
    \multicolumn{2}{l}{\it Orbital periods based on:}\\
    \hspace{2cm} Solar system moon periods &	14&54&24&92\\
    \hspace{2cm} \cite{cilibrasi2020nbody} exomoon models&	8&31&14&53\\
    \hspace{2cm} \cite{2019AJ....158...75H} M-dwarf exoplanets&	22&84&37&143\\
    \hline
    \multicolumn{2}{l}{\it Occurrence rates based on:}\\	
    \hspace{2cm} \cite{cilibrasi2020nbody} exomoon models &	14&54&24&92\\
    \hspace{2cm} \cite{2019AJ....158...75H} M-dwarf exoplanets&	7&47&11&65\\
    \hspace{2cm} One $5\times10^{-5} \times M_{\rm Host}$ satellite/system&	14&71&17&102    
    \end{tabular}
    \label{SummaryOfYields}
\end{table*}   

\subsection{Yields for Exoplanets Transiting Stars}
\subsubsection{Young Stellar Exoplanets}
Assuming the stellar IMF of \cite{2020ApJ...896...80G}, we expect the TEMPO survey to monitor approximately 2,000 young ONC stars,
with most exoplanet detections being associated with very-low-mass stellar hosts.
Therefore, we focus on the occurrence rates for M-dwarf hosts. 
These very-low-mass stars have extended radii at young ages, significantly increasing transit probabilities compared with main-sequence mid-to-late M dwarfs (see Figure~\ref{TransProbWAge}).

For very-low-mass stars, we calculated the expected number of detected transits based on the \cite{2015ApJ...807...45D} exoplanet occurrence rates and the \cite{2018AJ....155...89P} orbital periods.
For masses lower than those probed by past occurrence rate investigations, we assume a flat occurrence rate distribution anchored to the occurrence rate of the lowest-mass bin reported. 
In our calculation, we consider only the occurrence rate for orbital periods $<15$\,d  under the assumption that at least two transits will be required for a firm detection. 
We also assume that all exoplanets have H/He envelopes. For exoplanets $<6$\,\ME{}, we assume the envelope fractions calculated previously for an exosatellite embedded in a disk for 1.0\,Myr. For exoplanets $>6$\,\ME{}, we assume the mass-radius relation given in \cite{2017A&A...608A..72M} at 3\,Myr.

When calculating the occurrence rate of exosatellites transiting FFPs and BDs, we scale the satellite mass by the host mass. For example, when using solar system occurrence rates, we divide the moon mass by the host planet mass and use the same mass ratio for the simulated system. For exoplanets we use instead the exact exoplanet radius (without H/He envelopes) given in the distribution provided by \cite{2015ApJ...807...45D}. 
The \cite{2015ApJ...807...45D} investigation focused on early M dwarfs, which lies in the middle of our ONC stellar sample, motivating this assumption. 
Note that there are no Jupiter-mass planets in this distribution. 

The largest planets with available occurrence rates are 4\,$R_\Earth$. At 1--3\,Myr, these planets would contain H/He envelopes, making them about the same radius as Jupiter but far less massive \citep{2016ApJ...817..107O,2020MNRAS.498.5030O,2021arXiv211009531M}. 
Our approach for calculating transit yields likely overestimates the number of massive planets around late M dwarfs, while underestimating the number of massive planets around FGK stars. The approach also likely underestimates the number of planets in mid to late M-dwarf systems, which are known to host more planets than higher mass stars \citep{2019AJ....158...75H}.
An estimate of the number of detections around mid and late M dwarfs using \cite{2019AJ....158...75H} is in reasonably good agreement with \cite{2015ApJ...807...45D}, which suggests our yield numbers are plausible.

When determining our detection limits, we assumed that the stellar population was distributed evenly through the nebula, with stellar extinction ranging between $A_{\rm F146} =$~0--8~mag. 
Our transit yield calculations were performed for the F146 band both with and without WFI defocus (i.e., without and with saturation limits), assuming a minimum detectable transit depth of either $0.1\%$ or $0.01\%$. For the simulations of detected TEMPO exoplanets transiting young stars, we adopted as a baseline a minimum detectable transit depth of $0.01\%$, instead of the $0.05\%$ limit used for FFP and BD satellites, considering that the detection of exoplanets around young stars has previously been performed with this precision \citep{2018AJ....155....4M}.

\begin{figure*}[htb]
\centering
\includegraphics[width=0.9\textwidth]{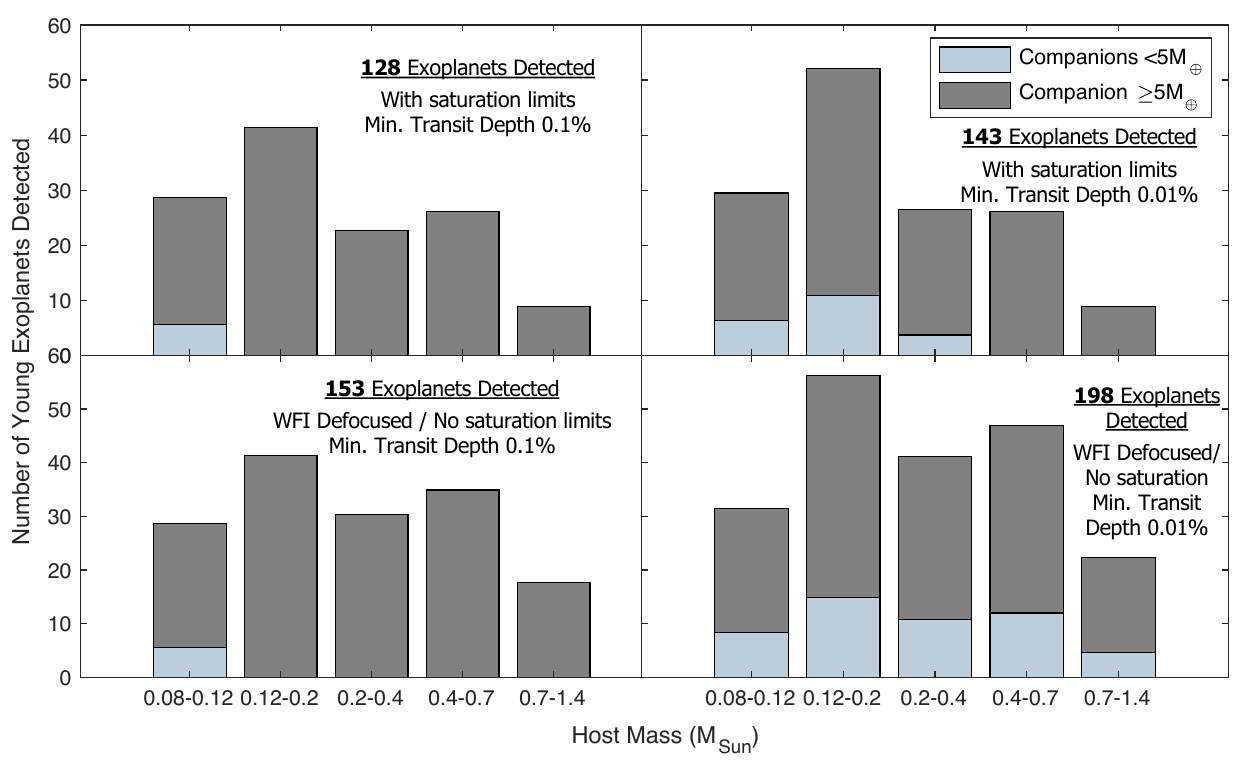}
\caption{Exoplanet yields around young (1--3\,Myr) stars in Orion observed with TEMPO. Yield calculations are performed for the WFI F146 band with (top two panels) and without (bottom two panels) instrument defocus (i.e., with and without saturation limits) and assuming a minimum detectable transit depth of $0.1\%$ (left two panels), and $0.01\%$ (right two panels). The use of WFI defocus and assuming a minimum detectable transit depth of $0.01\%$ optimizes the number of detected super-Titans.}
\label{YieldsExoplanets}
\end{figure*}

In Figure~\ref{YieldsExoplanets}, we illustrate our results. With WFI defocus and assuming a minimum detectable transit depth of $0.01\%$, there is a significant increase in the number of detected proto-terrestrial exoplanets (bottom right panel). 
This corresponds to the detection of 21 terrestrial ($<5.0$\,\ME{}) exoplanets and 122 exoplanets $>5.0$\,\ME{}.

There are no known analogs to these very young objects in the current exoplanet census. While some of these objects will likely be similar in mass to Earth, theory predicts that they should have \textit{significantly} larger radii than Earth due to the predicted presence of an H/He envelope. 
Probing their atmospheres will improve our understanding of the lower planet mass limit for envelope capture and loss. 
While we only expect the detection of about 50\% more exoplanets with WFI defocus, we emphasize that the additional exoplanets that will be detected are likely to be some of the most interesting (low mass proto-terrestrial) objects in this population. 

Further, incorporating the WFI defocus mode would allow the TEMPO survey to detect these additional transits around the brightest stars (i.e., those with little dust extinction). This stellar population is sufficiently bright (F146 = 12--15\,\ab{}) to permit follow-up transit spectroscopy, which may be possible on even the lower mass (Earth-mass) exoplanets due to the presence of an extended envelope (i.e., large atmospheric scale heights).  

\subsubsection{Field Star Exoplanets}
The TEMPO survey would monitor approximately 5,000 field stars for transits. 
To calculate detection yields for this population, we use the occurrence rates from \cite{2018AJ....155...89P} and the field star magnitude distribution from \cite{2020ApJ...896...79R}. 
When calculating transit probability we also assume a stellar radius of 1\,\Rsun{} for all stars. 
Assuming the WFI defocus mode, we expect ${\approx}$24 detected exoplanets with orbital periods $<30$\,d. 
We estimate that about 3 would have radii $<1.7$\,\RE{}, 13 would be Neptune-sized (1.8--5\,\RE{}), and eight Jupiter-sized ($>5$\,\RE{}). 
If instead, the WFI is not defocused, we expect to detect ${\approx}$17 exoplanets, of which 10 would be Neptune-sized and 7 Jupiter-sized. However, in that case, we do not expect to detect exoplanets $<1.7$\,\RE{}.

While there are many more field stars in the TEMPO FOV than young Orion stars, we expect to detect many more young transiting satellites and exoplanets than field star exoplanets.
This is because: (a) Orion stars are relatively bright, and spatially concentrated, making them ideal targets for efficient transit searches; (b) young FFPs, BDs, and stars have larger radii, thereby increasing the transit probability vs. their more evolved  counterparts; (c) young companions are likely to have H/He envelopes further increasing their radii and transit depths, making detection much easier; and (d) in contrast to field stars, the ONC stars we will monitor are primarily mid to late M dwarfs, most of which are expected to host multiple, short-orbit exoplanets \citep{2019AJ....158...75H}.

%%%%%%%%%%%%%%%%%%%%%%%%%%%%%%%%%%%%%%%%%%%%%%%%%%%%%%%%%%%%%%%
\section{Discussion}\label{Discussion}
\subsection{Filter Selection}
{\it Which band, F146 or F213, is expected to detect the most transiting exosatellites?} 
The analysis presented in the previous sections leads us to conclude that in the FFP regime, where we are always photon noise limited, TEMPO is likely to detect ${\sim}30\%$ more FFP moons/satellites in the F146 filter than in the F213 filter.
If TEMPO can detect transit depths $<0.05\%$, but is unable to address the detector saturation limits through WFI defocus, then the F213 filter offers some significant advantage above the FFP mass range, as it would detect more than double the number of BD satellites and low-mass star exoplanets vs. the F146 survey. However, this comes at the price of losing $30\%$ of the FFP satellites. 

If TEMPO is able to both detect transit depths $<0.05\%$ {\it and} address the detector saturation limits using WFI defocus mode, it would maximize the detection of both FFP/satellites and BD/low-mass star exoplanets with the F146 filter.
%This optimal scenario is illustrated by the solid gray line in Figure~\ref{NoiseLimits}. 
However, as discussed previously in Section~\ref{Saturation}, using the WFI defocus mode would seriously degrade the return of the survey for other, non-transit science cases.

\subsection{Uncertainties in the Expected Exosatellite Yields}
%Based on this calculation, for
Very little is known about the exomoon population \citep{2015ApJ...806...51H,2018AJ....155...36T}, nor is much known about young exoplanets akin to a primordial Earth.
Therefore, a census investigation would explore a critical new discovery space. 
The TEMPO survey is our first opportunity to probe the population of small companions to planetary-mass objects and BDs. 
While this represents a compelling argument for conducting such a survey,  it also underlines the difficulty of reliably predicting detection yields given the many unknowns in the parameter space to be probed. Our predictions for the TEMPO survey detection yields have a wide range of uncertainties due to a large number of poorly confined parameters.  
To summarize the assumptions of our applied methods:
\begin{itemize}
    \item We assume transits across the stellar diameter, which maximizes the S/N ratio. On average, however, the mean transit path is equivalent to an impact parameter of b~$\approx$~0.62.  On average, this decreases the S/N ratio by 21\% (but changes our yields by $<$5\% in all scenarios).
    \item We do not (and cannot) present a detailed study of the effect of substellar variability on the detectability of transits. Uninterrupted  long-term (30\,d) observations of substellar objects have never even been taken. So not can we not only use previous observations for injection-recovery tests but we also do not even fully know what kind of (possibly wavelength-dependent) effects we will see in these light curves and how they will affect transits.
    \item Detrending of long-term (30\,d) light curves of substellar objects from variability in search for transiting satellites has not been done previously. In our paper, we compute S/N ratios based only on photon noise and using a minimum transit depth cutoff to account for (sub)stellar variability. It is unclear, however, how well detrending algorithms, can actually remove variability prior to a transit search.
    \item We assume disk dispersal has occurred for all sources, however the disk dispersal time, especially for young BDs and FFPs, is poorly constrained and dispersion has most likely not occurred for all our sources.
\end{itemize}

\subsection{Potential Impact of the TEMPO Survey}
Regardless of whether the detection yields will turn out to be consistent calculations, it is clear that the knowledge gained by the TEMPO survey will have profound implications on our understanding of the formation and evolution of terrestrial worlds. The results of this investigation will allow differentiating between the existing exomoon formation models, and determine if the moons of the solar system's gas giants are similar to FFP exomoons. Given the numerous unknown variables discussed throughout this work, and the present lack of known objects in the parameter space TEMPO is going to explore, we emphasize that this is a realm of discovery where it is impossible to know what one will find. This is perhaps the most compelling motivation to conduct a TEMPO-like survey.

\begin{figure}[tbh]
\centering
\includegraphics[width=0.48\textwidth]{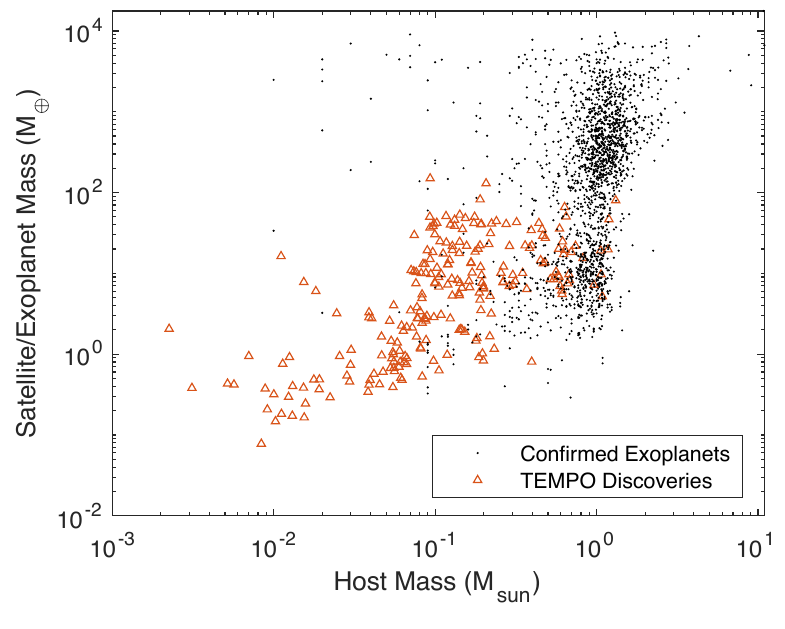}
\caption{Distribution of exosatellite or exoplanet mass (\ME{}) as a function of host mass (\Msun{}) for the simulated TEMPO survey detections (red triangles) and known exoplanet population (black dots). TEMPO discoveries will expand our understanding of exoplanet and exosatellite hosts from stellar down to FFP masses.}
\label{MvM}
\end{figure}

\begin{figure}[tbh]
\centering
\includegraphics[width=0.48\textwidth]{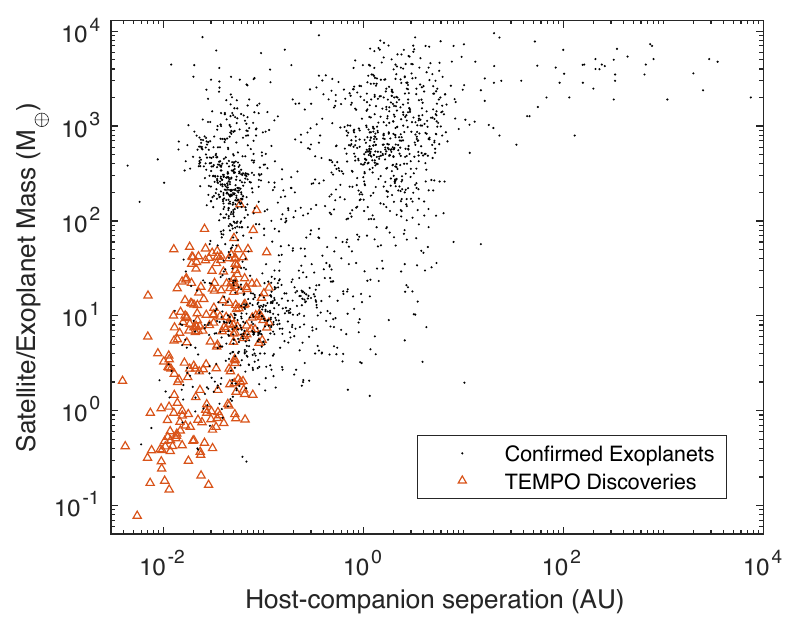}
\caption{Distribution of exosatellite or exoplanet mass (\ME{}) versus host-companion separation (AU) for the simulated TEMPO survey detections (red triangles) and known exoplanet population (black dots). TEMPO discoveries will expand our understanding exosatellites, including those down to separations comparable to the Galilean satellites (0.003-0.013\,AU).}
\label{MvAU}
\end{figure}

\begin{figure}[tbh]
\centering
\includegraphics[width=0.47\textwidth]{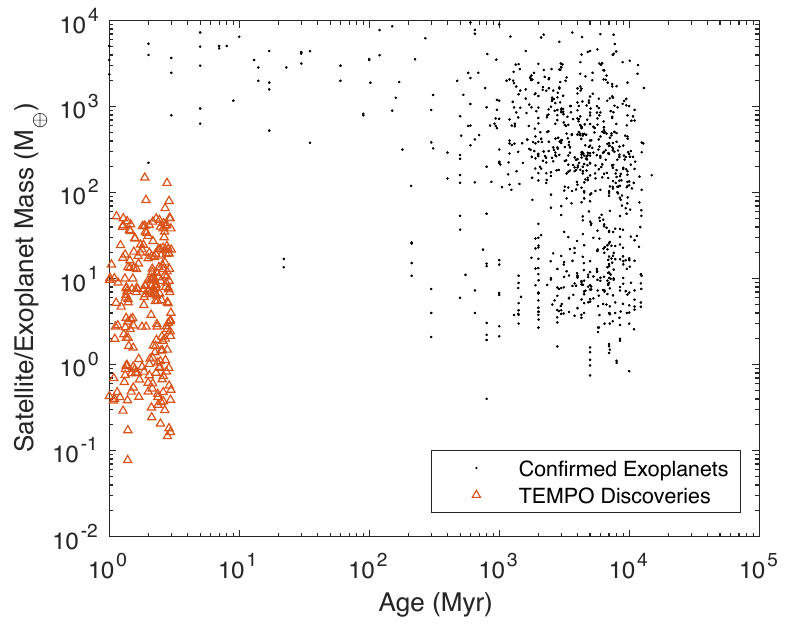}
\caption{Distribution of exosatellite or exoplanet mass (\ME{}) as a function of system age (Myr) for the simulated TEMPO survey detections (red triangles) and known exoplanet population (black dots). TEMPO discoveries will give us insight into exoplanet and exosatellite hosts at extremely young ages (1--3\,Myr).}
\label{MvAge}
\end{figure}

TEMPO will probe a completely new parameter space along the axes of host star mass and age. 
To illustrate the uniqueness of the young exoplanet and exomoon populations, we simulate the yield distributions of these systems. Using the default parameters from our yield simulations and without including the field star exoplanet detections,
we derive the counts shown in Table~\ref{tab:yields4TEMPO}.
We compare this population to the known detected exoplanet population.
In Figures \ref{MvM}-\ref{MvAge}, we illustrate the mass of the simulated companion detections (red triangles) from a TEMPO survey as a function of orbital separation (AU), system age (Myr), and host mass (\ME{}). This sample is compared to the population of known exoplanets (black dots). For the system age, we assume a uniform distribution of ages between 1--3\,Myr for TEMPO discoveries. Further results from this yield simulation are discussed in \cite{Limbach_Soares_InPrep2}, including a discussion of the fraction of companions one expects to detect in the ``proto-habitable zone." 

Only a handful of planets (or candidates) are known with ages less $<5$\,Myr and orbiting host stars with masses less than the approximately 80\,\MJ{} hydrogen-burning limit. Among them, none is potentially rocky, or low enough in mass to be the progenitors of rocky objects after losing the hydrogen envelope. TEMPO would not only detect additional examples of these elusive systems, but would potentially detect dozens of them, yielding a sample large enough for robust statistical work to study their demographics. Moreover, TEMPO would have enough sensitivity that detecting relatively few or even no transiting companion would significantly constrain current theories of satellite formation. 

\begin{table}[t!]
    \centering   
    \begin{tabular}{c|c}
    \multicolumn{2}{c}{Predicted TEMPO Detection Yields}\\
    \hline
    \hline
    \multirow{2}{*}{FFP Exosatellites}&12 Super-Titans ($\mathrm{M_{Titan}}$-\ME{})\\
    &2 Earths (1--2\,\ME{})\\
    \hline
    \multirow{2}{*}{BD Exosatellites}&34 Super-Titans ($\mathrm{M_{Titan}}$-\ME{})\\
    &20 Earths/Neptunes ($>1.0$\,\ME{})\\
    \hline
    \multirow{2}{*}{Stellar Exoplanets}&21 Proto-Terrestrial ($<5$\,\ME{}) \\
    &122 Exoplanets with $>5.0$\,\ME{}\\
    \hline
    \end{tabular}
    %\caption{}
    \label{tab:yields4TEMPO}
\end{table}

\subsection{Characterizing and Mitigating Host Variability}\label{HostVar}
%As we have discussed, the unexplored parameter space that the TEMPO survey will probe is significant, and yet another unknown is the degree of FFP and BD host variability at these extremely young ages. 
The atmospheres of FFPs and BDs are typically characterized by complex chemical processes that often lead to the formation of clouds. This can result in rotationally-modulated flux variability produced by cloud features rotating in and out of view \citep{Buenzli2014,Radigan2014,Metchev2015,Vos2022}. 
To first order, this rotational variability is sinusoidal, however, rapidly evolving variability  is also common. For example, the exomoon candidate transiting a 3.7\,\MJ,~10\,Myr FFP discussed in \cite{Limbach2021} exhibits a transit depth of $0.6\%$. The companion was marginally detected with a single transit in the presence of substantial FFP variability as well as significant photon noise. 
Young host variability is common not only in BDs and FFPs, but also in young stars. K2-33 and AU~Mic, for example, have shown evidence of flares in 3\,$\mathrm{\mu m}$ Spitzer data \citep{2016AJ....152...61M,2018ARep...62..532S,2022AJ....163..147G}. 
Disentangling young host variability from transit signals can be done utilizing a variety of existing techniques  \citep{2017AJ....154..224R,2017AJ....153...64M,2018AJ....155...10C,2018AJ....155....4M,2020AJ....160...33R,2021AJ....161...23M}. 

Transiting companion detection has been demonstrated for young stars down to 0.05\% transit depths \citep{2018AJ....156..266S,2018AJ....156...46V,2018AJ....155....4M}. However,
below a certain threshold value, variability can preclude the detection of some small companions. If variability represents the main source of noise,  it is difficult to predict the minimum detectable transit depth of the TEMPO survey, since our targets are less massive and younger than any transiting exoplanet host stars known to date.

Prior to Roman's launch, the best approach for determining how the host variability sets the noise limit would be a pilot study. 
A mid-sized JWST program could collect data on a portion of the TEMPO FOV for several days which could (1) provide constraints regarding the minimum detectable transit depth due to host variability; (2) provide the data necessary to determine which bandpass would produce the highest yield; (3) allow exploring the trade-off between grism and single-band observations (using simultaneous observations in both modes with NIRCam, which uses Teledyne detectors similar to Roman); and (4) provide data for the development of a pipeline that could be adapted and implemented for the TEMPO survey. 
Further, the JWST NIRCam FOV is sufficiently large that a 75-hr pilot program should detect at least one transiting exosatellite around either a BD or FFP in the ONC assuming our default model parameters listed in Table~\ref{YieldSettings} .

As a default for our yield calculations, we assumed a $0.05\%$ cutoff for the minimum detectable transit depth and we attributed this limit to the presence of host variability. 
We also provided yields using a 0.01\% and 0.1\% minimum detectable transit depth to probe how host variability may impact exosatellite detection. 
Nevertheless, we note that \cite{2018AJ....155....4M} detected a transit of $<0.02\%$ in the presence of young, stellar variability with a peak-to-peak amplitude more than $50\times$ the transit depth. 
Further, BDs seem to exhibit light curve variability morphologies that are similar in structure as young stellar exoplanet hosts \citep{Metchev2015,2020AJ....160...77Z,2021AJ....161..224T}. Assuming the young BDs in the ONC behave similarly, it stands to reason that the detection of transits down to 0.05\% should be possible with sufficient signal-to-noise.
Even if host variability may significantly reduce the number of detectable companions, particularly around BDs which are very inflated at young ages, it should not preclude the detection of small exosatellites including analogs to Titan and Ganymede transiting low-mass BDs and massive FFPs where detections are photon-noise (rather than host-variability) limited.

While variability from young stars, BDs, and FFPs can be difficult to distinguish from transiting companions in achromatic light curves, color information can break this degeneracy. 
This is because host variability is typically chromatic, whereas exosatellite transit signals are primarily gray. For example, numerous studies have found significantly lower variability amplitudes at mid-IR wavelengths compared to near-IR \citep{2020ApJ...903...15L,2020AJ....160...38V}
The spectral dependence of host variability amplitude has been studied in FFPs and BDs, but these studies have focused on objects older than those that TEMPO is going to observe in the ONC. The \cite{2020AJ....160...38V} study showed that the measured amplitude of L- and T-dwarfs in the J-band can vary from 0.5-26\%, with a median amplitude of 2.7\%, but the age of this population was 10\,Myr-1\,Gyr. Much of variability is sinusoidal and can be fitted and removed.
Further investigations  are required to characterize the spectral dependence of ONC host variability in Roman spectral bands. If the ONC FFP and BD hosts are significantly less variable in the F213 vs.~F146 band, then the TEMPO survey may be more successful if carried out in the redder (F213) Roman IR band.

Another option would be to switch between the F213 and F146 filters on short timescales (minutes). The host variability would differ in the two spectral bands, which would provide another means for disentangling exosatellite transits from host variability. Moreover, simultaneous multi-band observations would enable additional science investigations (e.g., stellar activity/flares/variability, dippers, dust extinction properties). The merits of this approach require further modeling to fully understand the trades between yield, duty-cycle, and the host-variability-induced detection limit. The impact of this observing mode on telescope operations and the lifetime of the mechanisms should also be taken into account.

Simultaneous multi-wavelength ground-based observations performed in tandem with the TEMPO survey would also expand the wavelength coverage, offering an opportunity to differentiate between host variability.
As there is only value added from simultaneous ground-based observations, one should supplement the TEMPO survey with simultaneous ground-based multiband time-series observations, leveraging telescopes such as the Vera Rubin Observatory, as well as time-resolved multi-object spectroscopy.
 Exploring and implementing multi-wavelength observations is particularly important in the high-mass BD and low-mass star range for ONC exosatellite transit searches, where host variability is expected to be the biggest obstacle.
BDs in the TEMPO FOV are sufficiently bright in the red optical (15--18 Vega mag in the $y$ band at 3\,Myr for 15-75\,\MJ{}) for Rubin to produce 
panchromatic light curves with sufficient photometric precision and angular resolution to permit the detection of 100\,ppm satellite transits.

\subsection{Euclid}
\label{euclidsec}
In this manuscript, we discussed the optimal design of a 30\,d ONC transiting satellite survey with Roman. Euclid is a visible-to-NIR space telescope that is under development by the European Space Agency, with a planned launch in 2023. 
Euclid's capabilities are comparable to Roman's in the following ways: (1) Euclid has a large telescope diameter (1.2~m, as compared to Roman's 2.4~m aperture); (2) Euclid is capable of visible, Y, J, and H band imaging and uses Teledyne detectors similar to those on Roman; and (3) Euclid has a large FOV (0.57\,deg$^2$; twice that of Roman).

Like Roman, Euclid is capable of searching for transiting exosatellites in the ONC. We modified the simulator we developed for Roman to estimate the yields we could expect from a 30\,d Euclid survey. We use the parameters given in Table~\ref{YieldSettings} with a few exceptions. To determine the S/N from Euclid observations, we scale Roman's photometric sensitivity based on the smaller Euclid aperture and narrower bandpass (Euclid's H-band filter). 

\begin{figure}[tbh]
\centering
\includegraphics[width=0.47\textwidth]{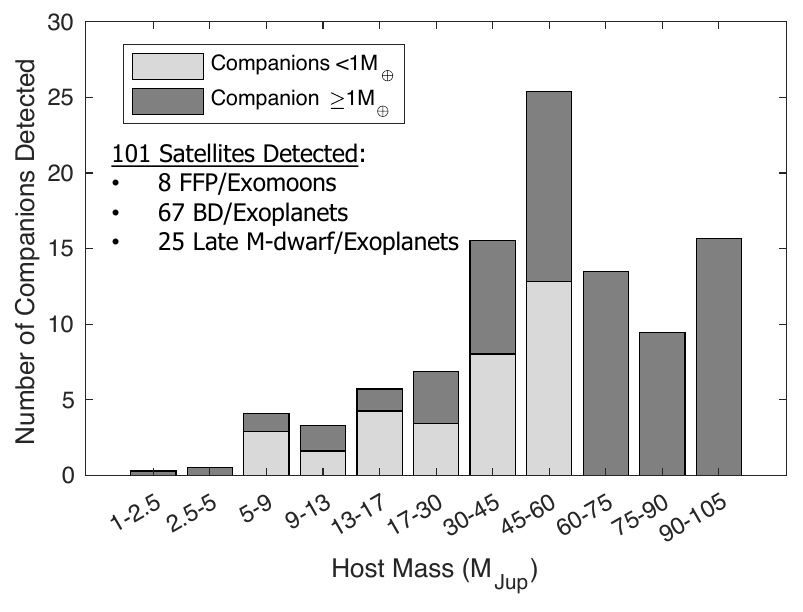}
\caption{Number of expected detections of transiting satellites with a 30\,d Euclid survey using the same baseline parameters as the 30\,d Roman survey (see Table~\ref{tab:para} for a review).
Super-Titans ($M_{\mathrm{Titan}}$-\ME{}) are shown in light gray, and satellites more massive than Earth are in dark gray. Euclid is capable of detecting more satellites transiting BDs (67 versus 54) than Roman due to the larger FOV. However, Euclid would detect fewer satellites transiting FFPs (14 versus 8) due to a lower photometric sensitivity.}
\label{Euclid}
\end{figure}

As shown in Figure~\ref{Euclid}, using our default parameters (Table~\ref{YieldSettings}) we estimate that a Euclid survey would detect eight exomoons transiting FFPs and 67 exosatellites transiting BDs. This can be compared to Roman's yield of 14 exomoons and 54 exosatellites.
Exomoon detection is typically S/N limited and is therefore more challenging with a smaller telescope.
Euclid is still capable of producing half the exomoon yield of Roman despite the smaller aperture. 
However, a Euclid survey would likely result in the detection of more BD exosatellites due to the larger FOV.
In summary, the synergy between the two facility would provide the first census demographics of the exomoon and exosatellite population, offering the scientific community the first insights into these populations.

\section{Summary}\label{summary}
In this paper, we predicted the detection yield of very young transiting companions in the ONC with a  30\,d TEMPO survey. Based on a large number of unknowns about this parameter space, we emphasized that it is difficult to identify and bound the inputs for these yield calculations and this is reflected in the wide range of yield predictions. 
We used a combination of various theoretical models and/or measured demographics of both M-dwarf exoplanets and solar system moons.
Using these assumptions, we found that the TEMPO survey would be able to detect a statistical sample of companions to planetary-mass objects and BDs. 
More specifically, we estimate the number of detections of BD or FFP satellites to range between 10 to more than 100 detections. 
Characterizing these populations  would allow us to differentiate between the demographic models employed in these simulations. 

The TEMPO survey offers a rare opportunity to explore the timescales and mechanisms for (a) exosatellite formation, (b) H/He envelope capture and loss, (c) companion migration, and (d) disk dispersal. 
In all the investigated scenarios, we determined the detection of exosatellites into the planetary-mass host range ($<13$\,\MJ{}).
For the chosen baseline parameters, we estimated a detection yield of a dozen ``moon-analogs" transiting FFPs and ${\approx}50$ exosatellites transiting BDs.
Detecting and characterizing this population would provide the first demographic study of a population of exosatellites orbiting sub-stellar objects and Jupiter-analogs, providing insight into the similarities between the detected exosatellites and the Galilean moons. 
In addition to the detection of BD and FFP exosatellites, we estimated that ${\approx}150$ young exoplanets would be detected in the TEMPO survey FOV. 
Many of these would be ``proto-habitable zone" worlds transiting low-mass stars, similar to an early-stage Trappist-1 system. 
These exceptionally young proto-habitable zone worlds would provide a window into terrestrial planets in their infancy. 

%\hfill \break
\section*{Acknowledgement}
We thank Enrico Ramirez-Ruiz and Darren L.~DePoy for their helpful conversations.
MAL acknowledges support from the George P. and Cynthia Woods Mitchell Institute for Fundamental Physics and Astronomy at Texas A\&M University. 
MSF gratefully acknowledges the generous support provided by NASA through Hubble Fellowship grant HST-HF2-51493.001-A awarded by the Space Telescope Science Institute, which is operated by the Association of Universities for Research in Astronomy, In., for NASA, under the contract NAS 5-26555. 
A.L.R. acknowledges support through the ITC Post-doctoral Fellowship funded by the Institute for Theory \& Computation at Harvard University and through the National Science Foundation (NSF) Astronomy and Astrophysics Postdoctoral Fellowship under award AST-2202249.
The Roman Space Telescope FOV footprint used in this work was obtained from the Mikulski Archive for Space Telescopes (MAST) at the Space Telescope Science Institute. STScI is operated by the Association of Universities for Research in Astronomy, Inc., under NASA contract NAS5–26555. Support to MAST for these data is provided by the NASA Office of Space Science via grant NAG5–7584 and by other grants and contracts.
This research has made use of the NASA Exoplanet Archive, which is operated by the California Institute of Technology, under contract with the National Aeronautics and Space Administration under the Exoplanet Exploration Program.

\facilities{Cycle 22 HST Treasury Program “The Orion Nebula Cluster as
a Paradigm of Star Formation” (GO-13826, P.I. M. Robberto; see \citealt{2020ApJ...896...79R}), Gaia DR3 \citep{GaiaDR3}, Mikulski Archive for Space Telescopes \citep{MAST}, Cycle 1 JWST “A Census to the Bottom of the IMF in Westerlund 2: Atmospheres, Disks, Accretion, and Demographics”	(GO-2640, P.I. William Best)}
\software{SAOImageDS9 \citep{ds9}}

\bibliography{main}{}
\bibliographystyle{aasjournal}
\end{document}